\documentclass{aa}

\usepackage{natbib}
\usepackage{graphicx}
\usepackage{txfonts}
\usepackage[colorlinks,allcolors=blue]{hyperref}

\usepackage{multirow}
\usepackage{xspace}
\usepackage{scalerel}
\usepackage[normalem]{ulem}

\newcommand{\cprime}{C$^{\prime}$}
\newcommand{\ks}{K_{\rm s}}
\newcommand{\wjk}{W_{\scaleto{\rm J,K_s}{4.5pt}}}
\newcommand{\wbprp}{W_{\scaleto{\rm BP,RP}{4.5pt}}}
\newcommand{\wvi}{W_{\scaleto{\rm V,I}{4.5pt}}}
\newcommand{\gbp}{G_{\scaleto{\rm BP}{4.5pt}}}
\newcommand{\grp}{G_{\scaleto{\rm RP}{4.5pt}}}
\newcommand{\gaia}{\textit{Gaia}}
\newcommand{\gtmd}{\gaia-2MASS diagram}

\begin{document}
  \title{Semi-regular red giants as distance indicators}
  \subtitle{I. The period--luminosity relations of semi-regular variables revisited\thanks{
    Table~\ref{tab:sample} is only available in electronic form
at the CDS via anonymous ftp to \url{cdsarc.u-strasbg.fr} (\url{130.79.128.5})
or via \url{http://cdsweb.u-strasbg.fr/cgi-bin/qcat?J/A+A/}
  }}
\titlerunning{PL relations of SRVs revisited}
\authorrunning{Trabucchi et al.}

\author{M.~Trabucchi\inst{\ref{inst_gen}}\fnmsep\thanks{Corresponding author: M. Trabucchi (\href{mailto:michele.trabucchi@unige.ch}{\tt michele.trabucchi@unige.ch})},
N.~Mowlavi\inst{\ref{inst_gen}},
T.~Lebzelter\inst{\ref{inst_vie}},
}
\institute{
Department of Astronomy, University of Geneva, Ch. Pegasi 51, CH-1290 Versoix, Switzerland\label{inst_gen}
\and 
University of Vienna, Department of Astrophysics, Tuerkenschanz\-strasse 17, A1180 Vienna, Austria\label{inst_vie}
}
\date{Received August 15, 2021; accepted September 9, 2021}

\abstract
{Semi-regular variables (SRVs) are similar to Miras in brightness, and they also follow one or more period-luminosity relations (PLRs), though not necessarily the same one as Miras. As potential standard candles they are more challenging than Miras because of their smaller variability amplitudes and less regular light curves, but they are substantially more numerous and especially promising for probing old stellar populations.}
{We aim to characterise the variability of SRVs, specifically focusing on their connection with Miras, in order to prepare the ground for investigating their potential as distance indicators.}
{We examine SRVs and Miras in the Magellanic Clouds from OGLE-III observations, with data from \gaia\ and 2MASS. After cleaning the sample of variability periods unrelated to pulsation, we classify each source by chemical type and combination of pulsation modes. We examine the results in terms of global photometric and pulsation properties.}
{We identify four SRV groups that fit the general evolutionary scenario predicted by theory. SRVs dominated by fundamental-mode pulsation are very similar to Miras, especially if mono-periodic. They further split into two subgroups, one of which follows the same sequence as Miras in the period--luminosity and period--amplitude diagrams, without discontinuity.}
{The similarities between Miras and SRVs suggest that the latter can be adopted as distance indicators in a way that is complementary to the use of the former, thereby at least doubling the available number of long-period variables (LPVs) suitable for use as distance indicators. The traditional amplitude-based separation between Miras and SRVs is not necessarily appropriate, and a more physically sound criterion should also involve pulsation periods. While this would require comparatively longer time-series, they are expected to become accessible in the coming years even for weak sources thanks to current and future large-scale surveys. The table of reclassified LPVs is made public.}


\maketitle

\section{Introduction}
\label{sec:Introduction}

Long-period variables (LPVs) are cool red giant stars representing the late evolutionary stages of low- and intermediate-mass stars. Their variability is due to stellar pulsation in multiple radial and non-radial modes \citep[e.g.][and references therein]{Wood_2015,Yu_etal_2020}, and therefore they follow several distinct period--luminosity relations \citep[PLRs; e.g.][]{Soszynski_etal_2007} which makes them promising standard candle candidates.

Mira variables are a subtype of LPVs characterised by the largest visual amplitudes, and are experiencing the latest stages of the asymptotic giant branch (AGB) evolutionary phase. They have recently been drawing attention for their potential as distance indicators \citep{Whitelock_2013,Yuan_etal_2017LMC,Yuan_etal_2017M33,Yuan_etal_2018,Huang_etal_2018,Huang_etal_2020}. Effort is being made to expand the available observational data sets, especially in the infrared \citep[IR;][]{Whitelock_etal_2008,Riebel_etal_2010,Ita_etal_2018} and with a focus on the longest periods \citep[with the potential to probe larger distances,][]{Karambelkar_etal_2019}, as well as to improve the detection and characterisation of their periods and PLRs \citep[e.g.][]{He_etal_2016,He_etal_2021}.

Miras are very bright (intrinsically and in the IR) and present in stellar populations of young and intermediate age. They display regular light curves with large visual amplitudes and follow a PLR \citep{Glass_LloydEvans_1981} that is especially well-defined at near-IR (NIR) wavelengths \citep{Glass_Feast_1982a}. In other words, they are common in various astrophysical environments, and are easy to observe and characterise \citep[e.g.][]{Whitelock_2019}.

Semi-regular (SR) variables are LPVs with shorter periods and smaller amplitudes than Miras, on average, and a lesser degree of regularity in their light curves. According to the General Catalogue of Variable Stars \citep[GCVS,][]{Samus_etal_2017_GCVS}, they can be distinguished from Miras by means of their visual amplitude of $\Delta V<2.5$ mag, and are split into semi-regulars of type a (SRa) and type b (SRb)
\footnote{
    We do not consider here semi-regulars of types c and d (SRc, SRd), consisting of red supergiants and yellow giants.
} (depending on whether or not their periodicity is well expressed). They are thought to be the progenitors of Miras, and they follow the same PLR, as well as a second relation at shorter periods \citep{Wood_Sebo_1996,Bedding_Zijlstra_1998}. Following the nomenclature of \citet{Wood_2015}, we indicate the Mira relation (or sequence) by the letter C, and use the label \cprime\ for the SR sequence.

One of the by-products of the advent of long-duration optical microlensing surveys (EROS, MOA, MACHO) is the detection of smaller-amplitude SR variables, and the discovery of additional PL sequences \citep{Wood_etal_1999,Wood_2000}. In particular, data gathered by the Optical Gravitational Lensing Experiment \citep[OGLE,][]{Udalski_etal_1992} allowed the identification of another candidate subtype of LPVs called OGLE Small Amplitude Red Giants \citep[OSARGs,][]{Wray_etal_2004}. No such type is present in the GCVS, and they would be traditionally identified as semi-regular or irregular variables\footnote{
    Slow irregular variable stars (type L in the GCVS) are often attributed this type when insufficiently studied.
}, if not perhaps even constant stars.

It is not clear whether or not, or how, the standard SRa/SRb classification overlaps with OSARGs and OGLE semi-regular variables (to which we refer hereafter as SRVs to avoid confusion with the `traditional' SR type). OSARGs are classified as such based on their collective properties as observed by OGLE, while the distinction between SRa and SRb is not clearly defined in the GCVS. Indeed, even the well-defined amplitude-based separation between Miras and SR red giants has been debated. Many SRa stars have light curves that are as regular as those of Miras, from which they differ only in amplitude, and so the traditional classification could be misleading \citep[e.g.][]{Kerschbaum_Hron_1992,Kiss_etal_2000,Lebzelter_Hinkle_2002}. However, there is some consensus that Miras pulsate only in the fundamental radial mode, while SR variables (including OSARGs) can be multi-mode pulsators. The exact modal assignment is still debated, but there is reasonable evidence \citep{Soszynski_etal_2013,Wood_2015,Trabucchi_etal_2017} that sequences \cprime\ and C should be attributed to radial pulsation in the first overtone mode (1OM) and fundamental mode (FM), respectively.

Semi-regular variables have potential as distance indicators in a comparable (or at least complementary) way to Miras, and have several advantages. They are significantly longer-lived than Miras, and are therefore more numerous in a given stellar population. Observations during the third phase of the OGLE project \citep[OGLE-III,][]{Udalski_etal_2008} showed that there are almost ten times more SRVs than Miras (almost 50 times if OSARGs are included) in the Magellanic Clouds \citep{Soszynski_etal_2009_LMC,Soszynski_etal_2011_SMC}.

From an observational standpoint, the fact that SR variables are less evolved than Miras facilitates their study. Indeed, they suffer shallower mass loss, and therefore their observed brightness is less biased by the presence of circumstellar material. This is important both in the IR \citep[where the emission from dust can become dominant, e.g.][]{Whitelock_etal_2017} and in the optical, where variability observations are normally carried out, and where excessive obscuration can render a star undetectable.

Semi-regular variables are especially promising as an alternative to RR Lyrae for probing old stellar populations as they are substantially brighter. They are also interesting for studying stellar systems showing extended star-formation history where other distance indicators are available. For instance, the optical brightness of SRs is comparable to that of classical Cepheids, compared to which they are normally brighter in the IR, and they can be more numerous \citep[e.g. in the Large Magellanic Cloud, based on OGLE-III data from][]{Soszynski_etal_2008}.
However, SR variables have so far received little interest in this sense compared with Miras \citep[e.g.][]{Tabur_etal_2010,Rau_etal_2019} as they are more difficult to deal with, the main obstacles being their smaller amplitudes and multi-periodicity.

This work is the first in a series of papers dedicated to investigating their potential in more detail, to verify the benefits they would provide, and to examine the requirements for their application. Here, we focus on preparing the ground by providing a general characterisation of sources observed by the OGLE program in the Magellanic Clouds. In the present, paper we do not study OSARG variables, but we focus on SRVs, which are the most similar to Miras in terms of luminosity and period range. The question of the effectiveness and practical use of SRVs as distance indicators will be addressed in a forthcoming paper.

Here we pursue the identification of subgroups of SRVs based on how many and which pulsation modes they exhibit. The interest in obtaining such a classification is two-fold: On the one hand, it provides a framework to better understand the loosely defined SRV class, and to more confidently place these stars in an evolutionary picture. On the other hand, it effectively ranks SRVs by how similar they are to Miras, which pulsate only in the fundamental radial mode.

For this purpose, we first have to identify and remove `contaminant periods', that is, periods that are not due to pulsation but are rather spurious or associated with other variability phenomena. We also have to obtain a reliable distinction between oxygen-rich and carbon-rich sources, as these show distinct variability patterns because of their different photometric properties and evolutionary history.

This paper is structured as follows. In Sect.~\ref{sec:Data} we present the adopted data set, while in Sect.~\ref{sec:Classification} we describe the process of cleaning and classification of the sample, first in terms of chemical type and therefore pulsation properties. Section~\ref{sec:AnalysisOfTheSample} is dedicated to the analysis of the resulting sample, focusing specifically on the SRV--Mira transition and on the systematic comparison of various SRV subgroups identified during the classification procedure. The results are interpreted and discussed in Sect.~\ref{sec:Discussion}, and we provide a summary and conclusions in Sect.~\ref{sec:Conclusions}.

\section{Data}
\label{sec:Data}

We adopt data from the OGLE-III catalogues of long-period variables in the Large Magellanic Cloud \citep[LMC,][]{Soszynski_etal_2009_LMC} and Small Magellanic Cloud \citep[SMC,][]{Soszynski_etal_2011_SMC}, consisting of mean $V$, $I$ magnitudes for 111 379 sources. These provide the three most prominent variability periods $P_k$ ($k=1,2,3$) derived by iteratively fitting and subtracting the third-order Fourier series corresponding to the main peak in the periodogram from the observed $I$-band light curve. The peak-to-peak amplitudes $\Delta I_k$ of the best-fit Fourier series for each period are also provided\footnote{
    When not referring to a specific one of the three periods or amplitudes, we often leave out the subscript and indicate them by $P$, $\Delta I$.
}.

It should be pointed out that only the primary period of each star has been visually verified \citep{Soszynski_etal_2009_LMC}, while secondary and tertiary periods were derived fully automatically. As irregular variability can cause the detection of false periodicities, this means that some secondary and tertiary periods from OGLE-III may be spurious. We also note that, since only three periods are reported for each source, the detection of spurious secondary and tertiary periods may cause true pulsation periods to be missed.

Besides variability information, the catalogues include a chemical-type classification of oxygen-rich (O-rich) and carbon-rich (C-rich) stars, as well as a distinction between variability subtypes, which are Miras, SRVs, and OSARGs. Both characterisations are based on photometric data. Details of the classification procedure are given by \citet{Soszynski_etal_2009_LMC}. Briefly, OSARGs were first singled out by their period ratios and position in the period--luminosity diagram (PLD). The remaining sources, consisting of both Miras and SRVs, were separated into O- and C-rich stars by their distribution in the diagram obtained with the Wesenheit indices,
\begin{equation}\label{eq:wvi}
    \wvi = I - 1.55\,(V-I)\,,
\end{equation}
\begin{equation}\label{eq:wjk}
    \wjk = \ks - 0.686\,(J-\ks)\,,
\end{equation}
where $V$, $I$ and $J$, $\ks$ are magnitudes in the OGLE and 2MASS filters, respectively.

To distinguish between Miras and SRVs, \citet{Soszynski_etal_2009_LMC} adopted a cut in $I$-band amplitude at $\Delta I=0.8$ mag, a choice justified by the sparsity of $V$-band time-series. They calibrated that criterion on the amplitude distribution of C-rich LPVs in their sample, whose bimodality they interpreted as the natural separation between SRVs and Miras. $V$-band amplitudes are not provided by OGLE, and so it is not easy to compare the adopted criterion with the traditional one. 

As pointed out by \citet{Soszynski_etal_2009_LMC} and clearly illustrated in their Fig.~3, the amplitude distribution of O-rich LPVs does not show a corresponding feature, and so the adopted SRV--Mira separation is arbitrary in this sense. Keeping in mind these remarks, in the following we rely on the variability types reported in the OGLE-III catalogue.

In the present study we focus on the 13 354 SRVs and 2015 Miras in the OGLE-III data set, and cross-match them with additional catalogues. We use near-infrared (NIR) photometry from the Two-Micron All-Sky Survey \citep[2MASS,][]{Skrutskie_etal_2006}, as well as optical photometry from the second data release and early third data release of the \gaia\ mission \citep[\gaia\ DR2 and EDR3,][]{GaiaCollaboration_2018_DR2,GaiaCollaboration_2020_EDR3}. For each OGLE source, we require an angular separation not larger than 1 arcsecond to the best-match counterpart from each other catalogue. This requirement reduces the size of the sample to 13 318 SRVs and 1931 Miras.

\section{Classification}
\label{sec:Classification}

\subsection{Chemical type identification}
\label{ssec:ChemicalTypeIdentification}
We employ \gaia\ photometry to corroborate the OGLE-III chemical-type classification. Following \citet{Lebzelter_etal_2018}, we construct the \gtmd\ (G2MD) with $\wjk$ (Eq.~\ref{eq:wjk}) and
\begin{equation}\label{eq:wbprp}
    \wbprp = \grp - 1.3\,\left(\gbp-\grp\right)\,,
\end{equation}
where $\gbp$ and $\grp$ indicate mean magnitudes in \gaia's blue and red photometers. While similar to the method adopted for the OGLE classification, it benefits from a wider optical spectral range for sampling molecular absorption features and characterising the atmospheric chemistry of LPVs \citep{Lebzelter_etal_2019}. \citet{Pastorelli_etal_2020} tested the G2MD against a sample of spectroscopically confirmed O- and C-rich asymptotic giant branch (AGB) stars in the LMC, and found a very high rate of success.

The \gaia\ scanning law often produces unevenly sampled time-series characterised by `clumps' of epochs very close in time \citep[with respect to the periods of LPVs; see e.g.][and references therein]{Mowlavi_etal_2018} that introduce a bias when the average magnitude is estimated. In principle, mean magnitudes from \gaia\ EDR3 should be more reliable as they benefit from longer observations, except this can also entail the introduction of additional epoch clumps in the light curve. This could be solved by reprocessing the light curves, properly accounting for data clumps, but LPV time-series are currently available only from DR2 \citep{Mowlavi_etal_2018}, and only for about 36\% of the OGLE-III Miras and SRVs in the Magellanic Clouds.

We therefore adopt an alternative approach. For each source, we estimate the chemical type with the G2MD computed with both DR2 and EDR3 mean photometry. If the results are not in agreement, we assume the mean magnitudes from one or both data releases are affected by the aforementioned issue, and cannot be used for the classification. If on the other hand the two methods are consistent with each other, we adopt the resulting chemical type (unless it is in disagreement with the type assigned by OGLE-III, in which case the source is also assigned uncertain chemical type).

This method cannot be applied to a number of sources that lack \gaia\ photometry in one or more filters (normally $\gbp$). Most of these are optically faint stars identified by OGLE as C-rich Miras, likely obscured by dust. Visual inspection of the $\wvi$~--~$\wjk$ diagram suggests that the OGLE-III classification is likely correct in most of these cases. Therefore, we adopt the OGLE-III classification for stars lacking \gaia\ photometry.

We obtain 6 376 C-rich SRVs, 6 510 O-rich SRVs, and 432 with uncertain identification, as well as 1 412 C-rich Miras, 462 O-rich Miras, and 57 with uncertain identification. We do not exclude unclassified sources, but only consider them in our analysis when no distinction is made by chemical type.

\subsection{Pulsation mode identification}
\label{ssec:PulsationModesIdentification}

\begin{figure}
    \centering
    \includegraphics[width=\hsize]{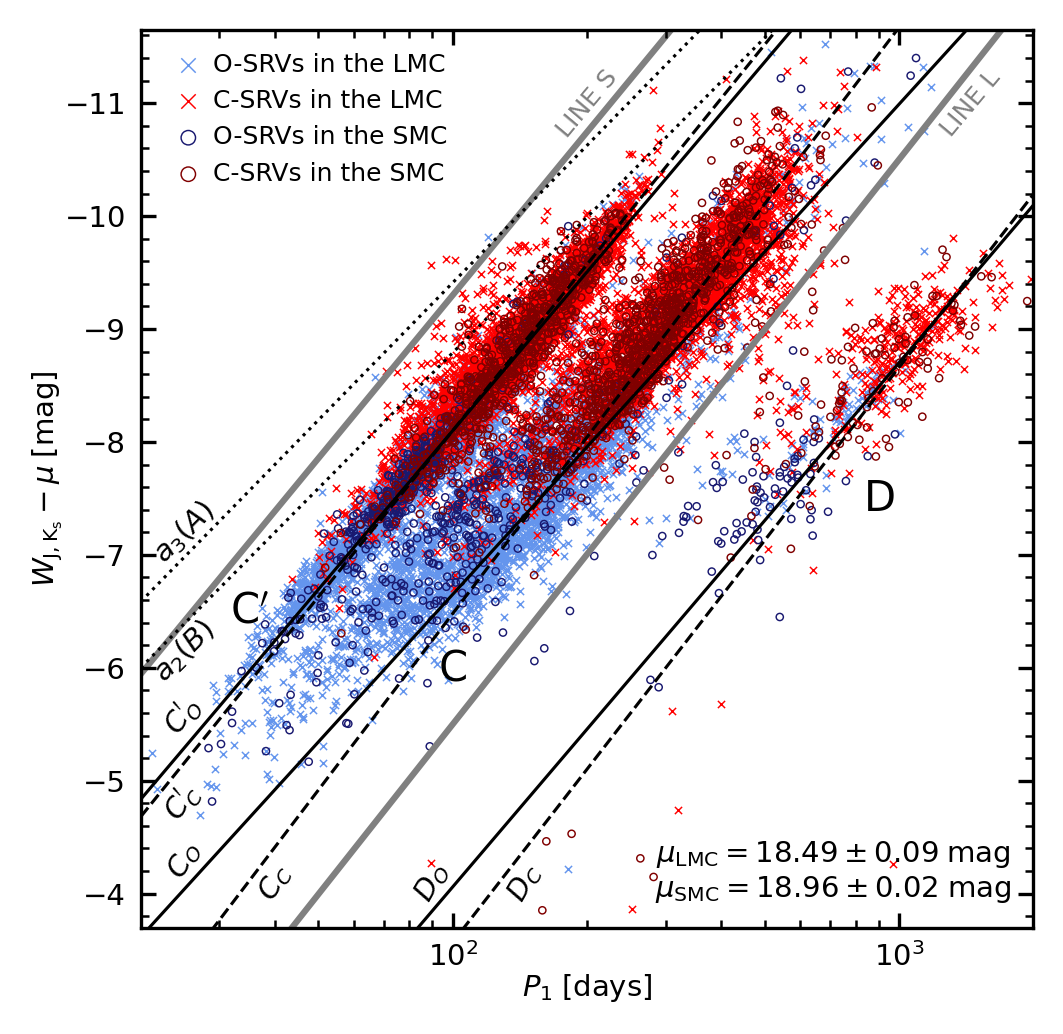}
    \caption{Primary \textit{variability} period versus absolute $\wjk$ for OGLE-III SRVs in the LMC (crosses) and SMC (circles), with tones of blue and red indicating O- and C-rich sources, respectively. Black lines are best fits to the PL sequences of LPVs in the LMC \citep[from][adjusted for a distance modulus $\mu=18.49$]{Soszynski_etal_2007}. More precisely, the best fits to sequences \cprime, C, and D are shown as solid or dashed lines depending on whether they were obtained on the sample of O- or C-rich stars. The dotted lines correspond to sequences A and B, which are referred to as $a_3$ and $a_2$ by \citet{Soszynski_etal_2007}. Thick grey lines are used to exclude periods from the sample (see text and Table~\ref{tab:pwjk_lines}).}
     \label{fig:PLD_SRVs_seqs}
\end{figure}

\begin{figure}
    \centering
    \includegraphics[width=\hsize]{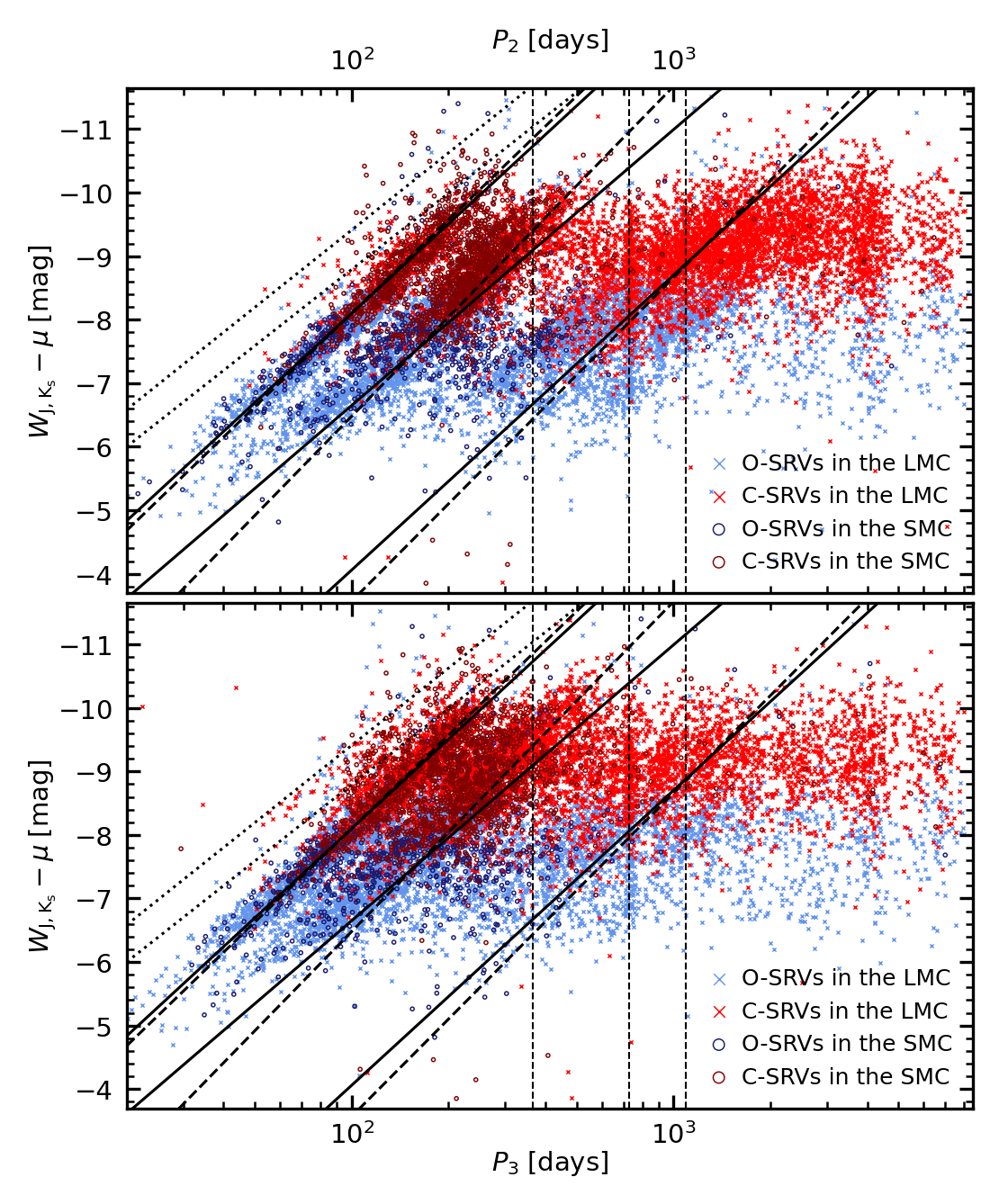}
    \caption{Similar to Fig.~\ref{fig:PLD_SRVs_seqs}, but for secondary and tertiary \textit{variability} periods. Vertical dashed lines mark periods of 1, 2, and 3 years to highlight the aliases due to seasonal gaps in the observations.}
     \label{fig:PLD_SRVs_seqs_P23}
\end{figure}

\begin{figure}
    \centering
    \includegraphics[width=\hsize]{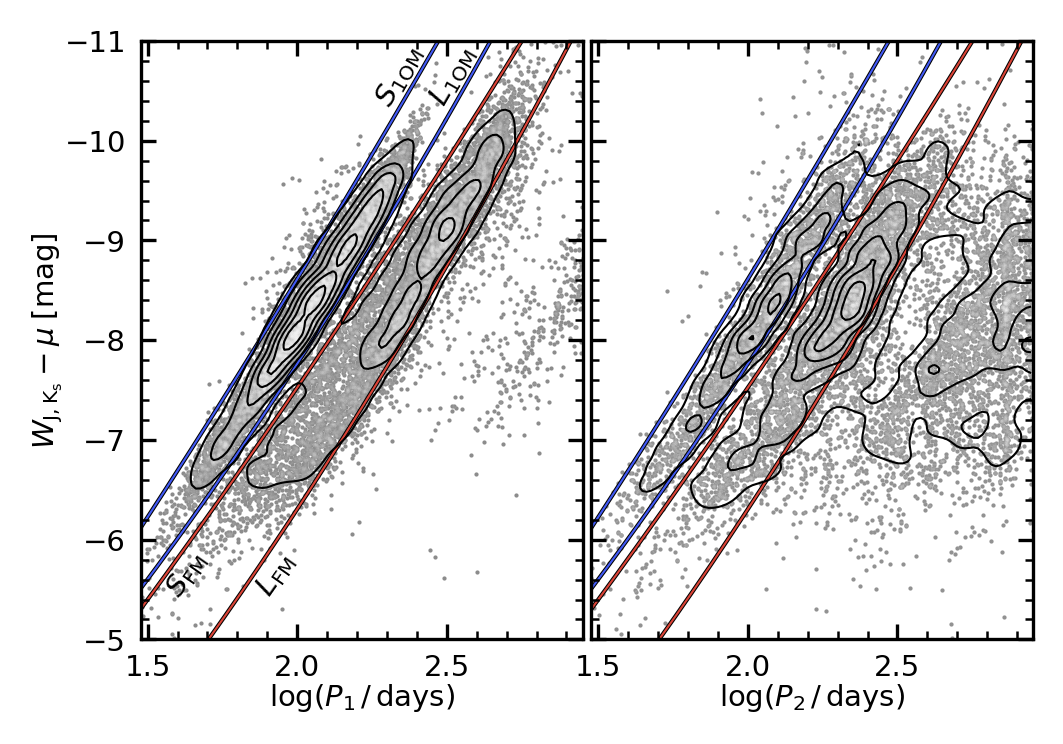}
    \caption{Period--luminosity diagram showing the boundaries adopted to assign the 1OM or FM to the periods of SRVs. Density contour lines are shown to visualise the slight offset of the secondary period sequences (right panel) towards longer periods compared to primary period sequences (left panel).}
     \label{fig:PLD_SRVs_seqOffset}
\end{figure}

\begin{table}
\caption{Coefficients of the equations defining boundary lines in the $\log(P)$ versus absolute $\wjk$ diagram (cf. Figs.~\ref{fig:PLD_SRVs_seqs} and~\ref{fig:PLD_SRVs_seqOffset}).}
\label{tab:pwjk_lines}
\centering
\begin{tabular}{p{0.1\columnwidth} p{0.1\columnwidth} p{0.1\columnwidth} p{0.1\columnwidth} p{0.1\columnwidth}}
\multicolumn{5}{c}{Line equation in the form:}\\
\multicolumn{5}{c}{$\wjk-\mu = a_0+a_1\,\log(\varepsilon\,P)+a_2\,[\log(\varepsilon\,P)-a_3]^2$}\\
\multicolumn{5}{c}{($\varepsilon=1$ for $P_1$, $\varepsilon=0.95$ for $P_{2,3}$)}\\
\hline
Line            & $a_0$& $a_1$& $a_2$& $a_3$\\
\hline
$S$             & 0.30 & -4.8 &      &      \\
$L$             & 4.50 & -5.0 &      &      \\
$S_{\rm 1OM}$   & 1.00 & -4.8 & -0.3 & 1.80 \\
$L_{\rm 1OM}$   & 1.85 & -4.8 & -0.6 & 2.15 \\
$S_{\rm FM}$    & 1.95 & -4.7 & -0.3 & 2.50 \\
$L_{\rm FM}$    & 3.10 & -4.7 & -0.6 & 2.10 \\
\end{tabular}
\end{table}

In the present section, we aim to refine the classification of the sample on the basis of variability properties, in particular in relation to multi-periodicity. We focus primarily on SRVs, while Miras, which are assumed to be single mode pulsators, are briefly discussed at the end of this section (in Sect.~\ref{sssec:Miras}).

The variability of LPVs is not only associated with stellar oscillations. In particular, LPVs often display so-called long secondary periods (LSPs) that form sequence D in the PLD and whose origin is still a matter of debate \citep[e.g.][]{Wood_2000,Olivier_Wood_2003,Wood_etal_2004,Derekas_etal_2006,Soszynski_2007,Wood_Nicholls_2009,Nicholls_etal_2009,Stothers_2010,Takayama_etal_2015,Saio_etal_2015,Takayama_Ita_2020,Pawlak_etal_2021,Soszynski_etal_2021}. As it is generally accepted that these periods are not due to pulsation, we aim to exclude them from the data set.

The sources in our selection have primary periods mainly on sequences \cprime\ and C, and less frequently on sequence D (Fig.~\ref{fig:PLD_SRVs_seqs}). Secondary and tertiary periods show the same pattern except they lie much more often in the long-period area of the PLD (Fig.~\ref{fig:PLD_SRVs_seqs_P23}). Not all of them are LSPs, as they can be spurious signals such as aliases resulting from seasonal observational gaps.

Sequences \cprime\ and C are compatible with pulsation in the 1OM (radial order $n=1$) and FM ($n=0$), respectively, according to the results of linear and nonlinear radial pulsation models \citep{Wood_2015,Trabucchi_etal_2017,Trabucchi_etal_2021}. On the basis of these considerations, our working hypothesis is that SRVs undergo radial pulsation either in the 1OM or FM, possibly at the same time\footnote{We note that our assumption concerns specifically the stars classified as SRVs in OGLE-III, as opposed to OSARGs. The latter, which would also be classified as semi-regular variables in the traditional scheme, likely pulsate in higher overtone modes as well, possibly non-radially.}, and they might also display LSP variability.

Therefore, we proceed first by identifying within the PLD the periods of SRVs that are unlikely to be caused by pulsation, which we then reject. We then identify the pulsation mode most likely responsible for each of the remaining periods.

\subsubsection{Identification of pulsation periods of SRVs}
\label{sssec:IdentificationOfPulsationPeriodsOfSRVs}

The main source of contamination in terms of variability not attributable to pulsation comes from the very long-period regime. Therefore, we exclude all primary periods longer than sequence C, which is those on the right side of line $L$ in the PLD (Fig.~\ref{fig:PLD_SRVs_seqs}). This is done in the $\log(P)$ versus absolute $\wjk$ plane \citep[with distance moduli $\mu_{\rm LMC}=18.49$, $\mu_{\rm SMC}=18.96$ from][]{deGrijs_etal_2017} in order to apply the same filter to both the LMC and SMC. Similarly, we reject short periods on the left side of line $S$. These are in the region of upper sequence A, attributed to the second overtone mode, and do not fit our hypothesis for the pulsation of SRVs.
In other words, we only retain primary periods between lines $S$ and $L$ (defined in Table~\ref{tab:pwjk_lines}) in the PLD. We apply the same filter for secondary and tertiary periods, except that before doing so we scale them down by a factor $\varepsilon=0.95$. As discussed later in this section, this scaling accounts for a small offset in the sequences formed by secondary and tertiary periods.

We emphasize that the process described above does not involve the rejection of catalogue sources, but only of their periods. A source is excluded only if all of its three periods are rejected. If, instead, only the primary period $P_1$ is rejected, it is replaced by $P_2$ (or by $P_3$ if $P_2$ is also rejected). Upon this filter and reorganisation, we indicate the remaining periods as $P_a$, $P_b$, and $P_c$ to emphasize the fact that we have now a high degree of confidence that they are actually {pulsation} periods, and to distinguish them from the more general {variability} periods $P_1$, $P_2$, and $P_3$ listed in the OGLE catalogues. We reject one period for 6 248 SRVs, two periods for 2 290 SRVs, and exclude 85 sources with no valid pulsation period.

\subsubsection{Pulsation mode assignment for SRVs}
\label{sssec:PulsationModeAssignmentForSRVs}

We assign a mode to each pulsation period by its position in the PLD. A period is identified with the 1OM if it is located on sequence \cprime\ (between lines $S_{\rm 1OM}$ and $L_{\rm 1OM}$, see Fig.~\ref{fig:PLD_SRVs_seqOffset}), and with the FM if it lies on sequence C (between lines $S_{\rm FM}$ and $L_{\rm FM}$).

These boundaries, whose expressions are given in Table~\ref{tab:pwjk_lines}, are defined on the distribution of primary pulsation periods. Upon applying them to the classification of secondary periods, we noticed that the latter display a small offset ($\simeq0.022$ dex), and the sequences they form are slightly displaced to the right with respect to primary periods. This can be appreciated in Fig.~\ref{fig:PLD_SRVs_seqOffset}, in particular for the periods on sequence \cprime. Tertiary periods display a very similar shift.

The origin of this offset, and whether it is of physical nature, is unclear. However, we verified that it can negatively impact our classification and the subsequent analysis. To account for this, a scaling factor $\varepsilon=0.95$ (defined by visual inspection) is applied to the value of secondary and tertiary periods before applying the classification defined for primary periods.

It is clear from Fig.~\ref{fig:PLD_SRVs_seqOffset} that further periods are excluded with this modal assignment. This helps to reduce the risk of contamination from spurious signals (`outside' the sequences, i.e. to the left of $S_{\rm 1OM}$ or to the right of $L_{\rm FM}$) or of incorrect mode assignment (for periods between the sequences, i.e. between $L_{\rm 1OM}$ and $S_{\rm FM}$), especially for secondary and tertiary periods.

Finally, for each star we verify that a given mode is assigned to no more than one period, which would be unphysical\footnote{
    This would be possible in the case of an oscillation with the same radial order and a different angular degree, which is not considered in our hypothesis. Instead we assume they are double detections of the same mode caused by period variations, which are common in SRVs.
}. In other words, a source cannot have more than one period on each sequence. As only two pulsation modes are considered here, this automatically leads to the exclusion of any tertiary period in each source. Then, if $P_b$ is on the same sequence as $P_a$, it is also excluded.

We identify 6 047 SRVs with two pulsation periods, 6 369 with a single pulsation period, and exclude 817 sources whose periods could not be clearly placed on either sequence \cprime\ or C. Table~\ref{tab:filtering} provides a summary of the SRV sample after the various filtering steps described in this section.

\begin{table}
\caption{Size of the SRV sample at each successive filtering step.}
\label{tab:filtering}
\centering
\begin{tabular}{c c c c c}
\multicolumn{2}{c}{Step} & LMC & SMC & Total \\
\hline
0   & OGLE-III SRVs                     &         11132 &           2222  &         13354  \\
\hline
\multirow{4}{*}{1}
    & $\theta_{\rm 2MASS}\leq1"$        &         11106  &          2215  &         13321  \\
    & $\theta_{Gaia}\leq1"$             &         11128  &          2218  &         13346  \\
    & all $\theta\leq1"$                & \textbf{11105} &  \textbf{2213} & \textbf{13318} \\
    & rejected                          &            27  &             9  &            36  \\
\hline
\multirow{5}{*}{2}
    & 3 periods                         &          2931  &          1764  &          4695  \\
    & 2 periods                         &          5948  &           300  &          6248  \\
    & 1 period                          &          2166  &           124  &          2290  \\
    & total                             & \textbf{11045} &  \textbf{2188} & \textbf{13223} \\
    & rejected                          &            60  &            25  &            85  \\
\hline
\multirow{6}{*}{3}
    & $P_a$ on \cprime, $P_b$ on C      &          2819  &           799  &          3618  \\
    & $P_a$ on C, $P_b$ on \cprime      &          1884  &           545  &          2429  \\
    & only $P_a$ on \cprime             &          3349  &           308  &          3657  \\
    & only $P_a$ on C                   &          2301  &           411  &          2712  \\
    & total                             & \textbf{10353} &  \textbf{2063} & \textbf{12416} \\
    & rejected                          &           692  &           125  &           817  \\
\hline
\multirow{3}{*}{3$^{\prime}$}
    & C-rich                            &          4647  &          1261  &          5908  \\
    & O-rich                            &          5464  &           645  &          6109  \\
    & uncertain                         &           242  &           157  &           399  \\
\end{tabular}
\tablefoot{The steps are: starting sample (0), well-matched sources (1), exclusion of non-pulsation periods (2), and mode assignment (3, corresponding to the final sample). The last few rows (3$^{\prime}$) give the numbers of sources in the final sample by chemical type, according to the approach described in Sect.~\ref{ssec:ChemicalTypeIdentification}. We note that all SRVs with $\theta_{Gaia\,{\rm DR2}}\leq1"$ also have $\theta_{Gaia\,{\rm EDR3}}\leq1"$, hence \gaia\ DR2 and EDR3 are not distinguished in the rows describing step (1).}
\end{table}

\subsubsection{Miras}
\label{sssec:Miras}

All but a few Miras have their primary variability period on sequence C. We identify seven Miras that do not fit this pattern. Secondary and tertiary periods of Miras display a chaotic distribution in the PLD, with no clear indication that they follow any PLR. This is consistent with the idea that Miras are mono-periodic\footnote{
    Since we focus on pulsation, throughout this paper we will use the term ``mono-periodic'' to indicate stars that display a single \textit{pulsation} period, as opposed to stars that pulsate in two or more modes simultaneously. This should not be interpreted as an absence of any other \textit{variability} period (such as an LSP), unless otherwise stated.}
, fundamental mode pulsators.

However, inspection of period ratios revealed that some Miras behave similarly to SRVs with a primary period on sequence C and a secondary period on sequence \cprime. Unfortunately the period ratio is often close to $P_2/P_1\simeq0.5$, and so it is difficult to determine whether the secondary periods are actually due to pulsation or are rather aliases of the primary period. While we do not exclude the possibility that some of these Miras are in fact pulsating in both the FM and the 1OM, here we reject all secondary and tertiary periods of Miras because of their uncertain origin.

For similar reasons, we do not attempt to use these secondary periods as a replacement for the primary periods in the few cases in which the latter do not lie along the PLR. We note that such displacement in the PLD can be due to an apparent dimming caused by circumstellar extinction, which is relatively common for Miras. In principle, this should be accounted for by the use of the approximately reddening-free Wesenheit index $\wjk$. When this is not the case, a convenient choice is to construct the PLD with the magnitude in some MIR filter, where the effect of extinction from circumstellar dust is significantly reduced. We adopted this approach with the help of photometric data from the all-sky catalogue of the Wide-field Infrared Survey Explorer \citep[AllWISE,][]{Cutri_etal_2003}. We checked the position of the seven outlying Miras in the AllWise PLDs and found no evidence that their displacement from the PLRs can be clearly attributed to dust extinction, so we decided to exclude them from the sample. We also exclude the source \textsc{OGLE-LMC-LPV-12184}, which is the Galactic Mira RX Dor.

The final sample of Miras consists of 1 923 sources, all of which have a single pulsation period that coincides with the primary variability period (i.e. $P_a=P_1$).

\subsection{Summary}
\label{ssec:Summary}

\begin{table*}
\caption{Sample lines of the data set with revised period classification (full table available at the CDS).}
\label{tab:sample}
\centering
\begin{tabular}{c c c c c c c c c c c}
OGLE-III ID$^(a)$ & $P_a$ & $P_b$ & $\Delta I_a$ & $\Delta I_b$ & $n_a^(b)$ & $n_b^(b)$ & $k_a^(c)$ & $k_b^(c)$ & Group & Ch. Type \\
          & [days] & [days] &        [mag] &        [mag] &       &       &       &       &       &          \\
\hline
OGLE-LMC-LPV-00017 &  82.04 &        & 0.097 &       & 1 &   & 1 &   &   SRV-1 & O \\
OGLE-LMC-LPV-00020 & 271.3  &        & 0.243 &       & 0 &   & 1 &   &   SRV-0 & C \\
OGLE-LMC-LPV-00034 & 158.4  &  89.57 & 0.103 & 0.065 & 0 & 1 & 1 & 3 & SRV-0.1 & O \\
OGLE-LMC-LPV-00035 &  95.62 & 157.43 & 0.085 & 0.074 & 1 & 0 & 1 & 2 & SRV-1.0 & O \\
OGLE-LMC-LPV-00038 &  67.81 & 135.08 & 0.031 & 0.036 & 1 & 0 & 1 & 2 & SRV-1.0 & O \\
                   ... & ... & ... & ... & ... & ... & ... & ... & ... & ... & ... \\
OGLE-SMC-LPV-00006 &  71.94 &        & 0.044 &       & 1 &   & 1 &   &   SRV-1 & O \\
OGLE-SMC-LPV-00015 & 269.3  &        & 1.014 &       & 0 &   & 1 &   &    Mira & C \\
OGLE-SMC-LPV-00022 & 282.0  &        & 1.397 &       & 0 &   & 1 &   &    Mira & C \\
OGLE-SMC-LPV-00023 & 188.26 &        & 0.600 &       & 0 &   & 1 &   &   SRV-0 & C \\
OGLE-SMC-LPV-00024 & 113.71 &  80.37 & 0.038 & 0.036 & 0 & 1 & 2 & 3 & SRV-0.1 & U \\
                   ... & ... & ... & ... & ... & ... & ... & ... & ... & ... & ... \\
\end{tabular}
\tablefoot{
$^{(a)}$ OGLE-III identifier as in \citet{Soszynski_etal_2009_LMC,Soszynski_etal_2011_SMC};
$^{(b)}$ radial order of pulsation assigned to the corresponding period ($n=0$ for the FM, $n=1$ for the 1OM); $^{(c)}$ rank assigned to the corresponding period in the OGLE-III catalogue, where $k=1$, $2$ or $3$ means that it is reported, respectively, as primary, secondary, or tertiary period by \citet{Soszynski_etal_2009_LMC,Soszynski_etal_2011_SMC}.
 }
\end{table*}

\begin{table*}
\caption{
Number of sources in the final sample according to the classification described in Sect.~\ref{sec:Classification}.}
\label{tab:puls_chem}
\centering
\begin{tabular}{r | c c c c | c c c c | c c c c}
\multirow{2}{*}{Group$^{(a)}$} & \multicolumn{4}{c|}{LMC} & \multicolumn{4}{c|}{SMC} & \multicolumn{4}{c}{Total} \\
 & O-rich & C-rich & unc.$^{(b)}$ & total & O-rich & C-rich & unc.$^{(b)}$ & total & O-rich & C-rich & unc.$^{(b)}$ & total \\
 \hline
 SRV-1   &  1520 &  1745 &    84 &  3349 &   121 &  156 &    31 &   308 &  1641 &  1901 &   115 &  3657 \\
 SRV-1.0 &  1620 &  1133 &    66 &  2819 &   237 &  512 &    50 &   799 &  1857 &  1645 &   116 &  3618 \\
 SRV-0.1 &  1115 &   736 &    33 &  1884 &   126 &  378 &    41 &   545 &  1241 &  1114 &    74 &  2429 \\
 SRV-0   &  1209 &  1033 &    59 &  2301 &   161 &  215 &    35 &   411 &  1370 &  1248 &    94 &  2712 \\
 \hline
 SRVs    &  5464 &  4647 &   242 & 10353 &   645 & 1261 &   157 &  2063 &  6109 &  5908 &   385 & 12416 \\
 (O- \& C-rich)  & \multicolumn{2}{c}{10111}  & \_ & \_ &\multicolumn{2}{c}{1906}& \_ & \_ &\multicolumn{2}{c}{12017}& \_ & \_ \\
 \hline
 \hline
 Miras   &   428 &  1106 &    53 &  1587 &    33 &  300 &     3 &   336 &   461 &  1406 &    56 &  1923 \\
 (O- \& C-rich) & \multicolumn{2}{c}{1534}   & \_ & \_ &\multicolumn{2}{c}{333} & \_ & \_ &\multicolumn{2}{c}{1867} & \_ & \_ \\
 \hline
 \hline
 SRVs \& Miras &  & & & 11940 &  & & & 2399 &  & & & 14339 \\
\end{tabular}
\tablefoot{
$^{(a)}$ the ``Group'' column distinguishes between Miras and SRVs (according to OGLE-III), and for the latter distinguishes between sources that are mono-periodic and pulsate in the FM (SRV-0) or 1OM (SRV-1), or that pulsate in both modes and predominantly in the FM (SRV-0.1) or 1OM (SRV-1.0);
$^{(b)}$ uncertain chemical type (Sect.~\ref{ssec:ChemicalTypeIdentification}).}
\end{table*}

\begin{figure}
    \centering
    \includegraphics[width=\hsize]{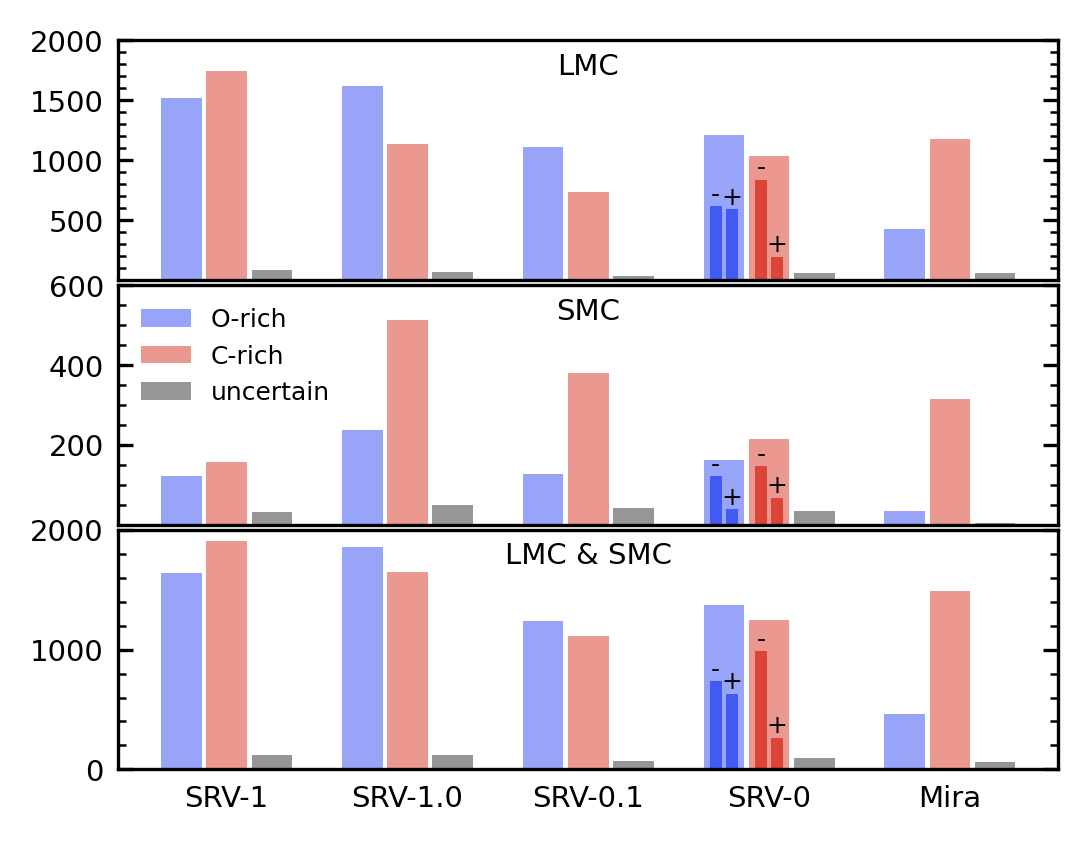}
    \caption{Histograms for the data reported in Table~\ref{tab:puls_chem}. Panels from top to bottom describe the LMC, SMC, and both. O- and C-rich sources are indicated by blue and red bars, respectively, while black bars correspond to sources whose chemical type is uncertain. Thin bars in the SRV-0 histograms indicate the subgroups SRV-0$^-$ and SRV-0$^+$, which are introduced in Sect.~\ref{ssec:TheSRVMiraTransition}.}
     \label{fig:hist_table_numbers}
\end{figure}

The final list of SRVs and Miras together with their pulsation parameters and the results of the classification by chemical type and variability properties are presented in Table~\ref{tab:sample} and made public through the CDS.
Table~\ref{tab:puls_chem} and Fig.~\ref{fig:hist_table_numbers} provide an overview of the classification results in terms of numbers of sources, both in total as well as for the LMC and SMC individually, and are useful to summarise the outcome of the procedure presented in Sect.~\ref{sec:Classification}.

O-rich SRVs are more abundant than C-rich ones in the LMC, while the ratio is strongly in favour of C-rich SRVs in the SMC. The chemical type is considered uncertain only for a small fraction of sources. In the LMC, 40\% of these sources would be C-rich according to the OGLE-III classification, while this fraction is 66\% for the SMC. Generally speaking, these numbers are consistent with the fact that LPVs in the SMC are relatively metal-poor compared with the ones in the LMC, hence the production of C-stars is favoured.

Let us now consider the proportions of SRVs by the number of periods and pulsation mode they have been assigned. Roughly half of the sample consists of SRVs with a single pulsation period; if it is the 1OM then we indicate these by SRV-1, otherwise they are marked SRV-0. The latter group is overall smaller in size among LMC sources, while in the SMC it is slightly more numerous than the SRV-1 group.

\begin{figure*}
    \centering
    \includegraphics[width=\hsize]{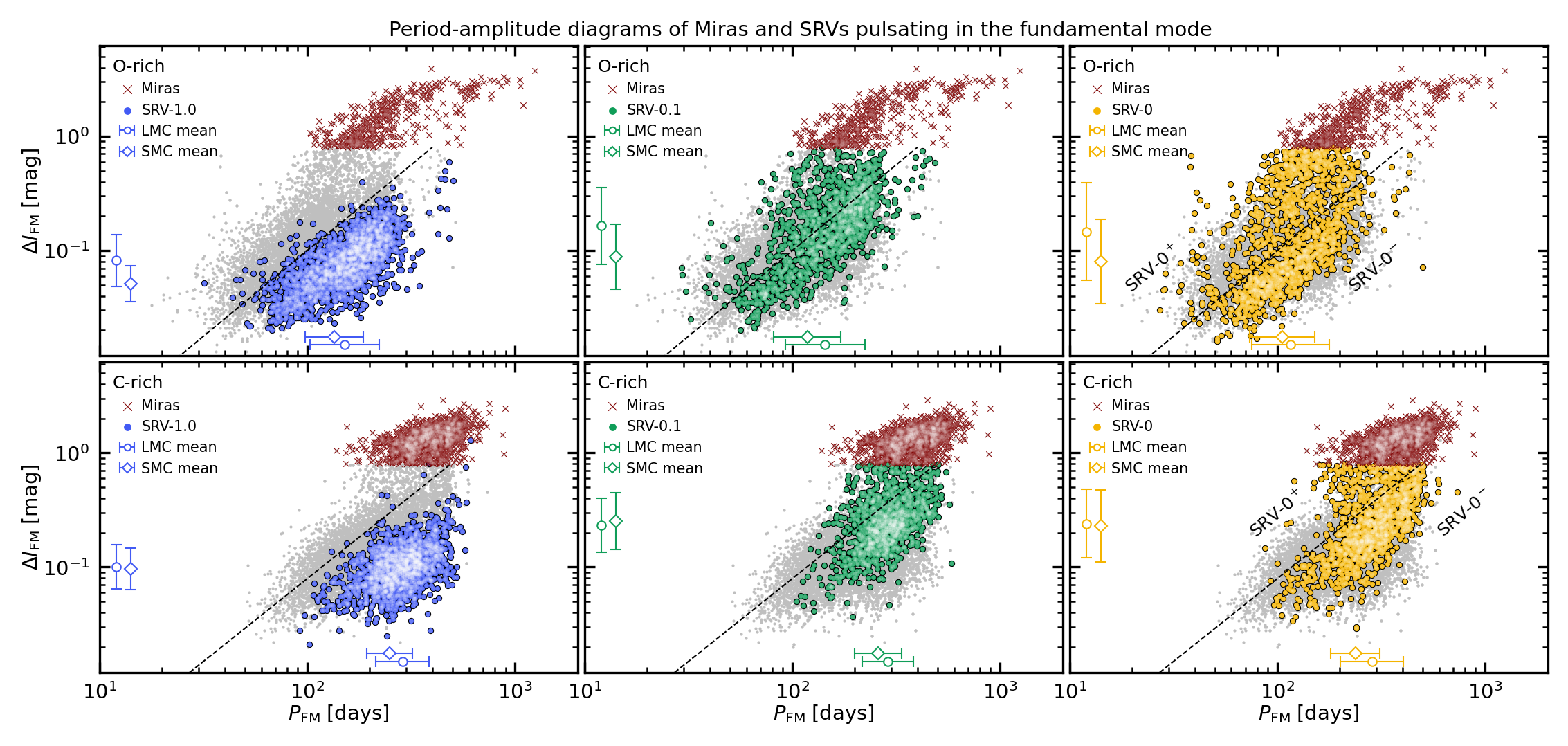}
    \caption{Fundamental mode period--amplitude diagrams of SRVs (circles) and Miras (crosses). The former are colour-coded according to the group they are assigned to, i.e. SRV-1.0 (blue), SRV-0.1 (green), and SRV-0 (orange), which are displayed in panels on the left, central, and right columns, respectively. Grey dots in the background show all the periods in the sample for visual reference. Symbols near the sides of each panel indicate the mean period and amplitude for each group, distinguishing between the LMC (empty circles) and SMC (empty diamonds), while error bars trace the corresponding standard deviation (these statistics are computed on a logarithmic scale). Panels in the top and bottom row are limited, respectively, to O- and C-rich LPVs. Dashed lines are defined by Eq.~\ref{eq:SRV0split}.}
     \label{fig:PAD_FM_SRVMira}
\end{figure*}

The remaining SRVs display two pulsation periods (i.e. they have a period both on sequence \cprime\ and on sequence C). Sources whose primary pulsation period ($P_a$) is identified with the 1OM are labelled SRV-1.0, otherwise the label SRV-0.1 is adopted. The latter group contains fewer sources than the former.

It should be kept in mind that the number of detected periods in the light curve of a LPV is not necessarily informative of its physics, as this number depends on instrumental and observational constraints. For instance, it is more difficult to extract secondary periods from a time-series that is relatively short or poorly sampled. Therefore, we checked the duration and number of epochs of the OGLE-III time-series and found no significant difference between sources with one or two pulsation periods. Time-series normally cover around 15-20 FM cycles and around 30 1OM cycles, and have between 300 and 1200 epochs, which corresponds to a mean of 40 epochs per cycle of the longest period. This suggests that the sample is sufficiently homogeneous for a classification based on the number of detected periods to be meaningful.

Moreover, all selected SRVs and Miras have pulsation amplitudes of larger than $\sim0.01$ mag, which is about one order of magnitude larger than the photometric uncertainty of OGLE. Therefore, we can consider the sample as complete from the point of view of variability.

Nonetheless, it is likely that at least some sources that we classify as mono-periodic are actually multi-periodic and have a secondary pulsation period that went undetected, possibly missed because of the detection of spurious secondary and tertiary periods (as discussed in Sect.~\ref{sec:Data}). As this bias is caused by a lack of information, it is not possible to estimate the fraction of sources affected by it. However, we can reasonably expect their number to be small.

Highly obscured dusty AGB stars represent a source of bias as they are unlikely to be observed in the OGLE $I$ band, or to have a valid cross-match with \gaia. However, these stars are relatively rare as they are typically massive, highly evolved objects. Moreover, they hardly fit the NIR PLR, and are therefore not useful in this sense as distance indicators.

The most important source of bias is perhaps associated with optically bright LPVs that are saturated in OGLE. Most of these stars correspond to red supergiants (RSGs) and high-mass O-rich AGB stars, i.e. branch $d$ of the G2MD, which is scarcely populated by OGLE-III sources compared with \gaia\ DR2 LPVs \citep[cf.][]{Lebzelter_etal_2018}. However, we note that the latter catalogue is affected by an amplitude bias, and so it is not easy to assess completeness in this sense. Branch $c$ stars (intermediate-mass O-rich AGBs) are similarly affected, but to a minor extent. These sources are discussed in more detail in Sects.~\ref{ssec:ComparisonBetweenSRVGroups} and~\ref{ssec:ColourMagnitudeDiagrams}.

\section{Analysis of the sample}
\label{sec:AnalysisOfTheSample}

\subsection{The SRV--Mira transition}
\label{ssec:TheSRVMiraTransition}
The period--amplitude diagram (PAD) is a powerful tool to investigate the observed variability properties of pulsating stars. Figure~\ref{fig:PAD_FM_SRVMira} shows the PADs of FM-pulsating LPVs, including Miras, separated by chemical type. As long as the FM is not dominant (SRV-1.0), the amplitude ranges occupied by SRVs and Miras are clearly separated, while this is not true for FM-dominated SRVs. In particular, SRVs showing only FM pulsation (SRV-0, right-column panels of Fig.~\ref{fig:PAD_FM_SRVMira}) appear to split into two subgroups. To further examine this feature, we adopt a tentative separation between the two SRV-0 subgroups defined by
\begin{equation}\label{eq:SRV0split}
    \log(\Delta I_a) = m\,\log(P_a)-4 \,,
\end{equation}
(where $m=1.50$ or $1.45$ for O- and C-rich sources respectively), which is displayed as a dashed line in Fig.~\ref{fig:PAD_FM_SRVMira}. The sources above and below the line given by Eq.~\ref{eq:SRV0split} are indicated
as SRV-0$^+$ and SRV-0$^-$, respectively. This splits the O-rich group of SRV-0 (633 are SRV-0$^+$ and 737 are SRV-0$^-$) into roughly equal parts, while the SRV-0$^-$ sources are predominant in the C-rich case (985 against 263).

We note that, compared with the classification based on pulsation periods, the distinction between SRV-0$^+$ and SRV-0$^-$ stars is arbitrarily defined for the purpose of further analysis, and should only be taken as a reference.

\begin{figure}
    \centering
    \includegraphics[width=\hsize]{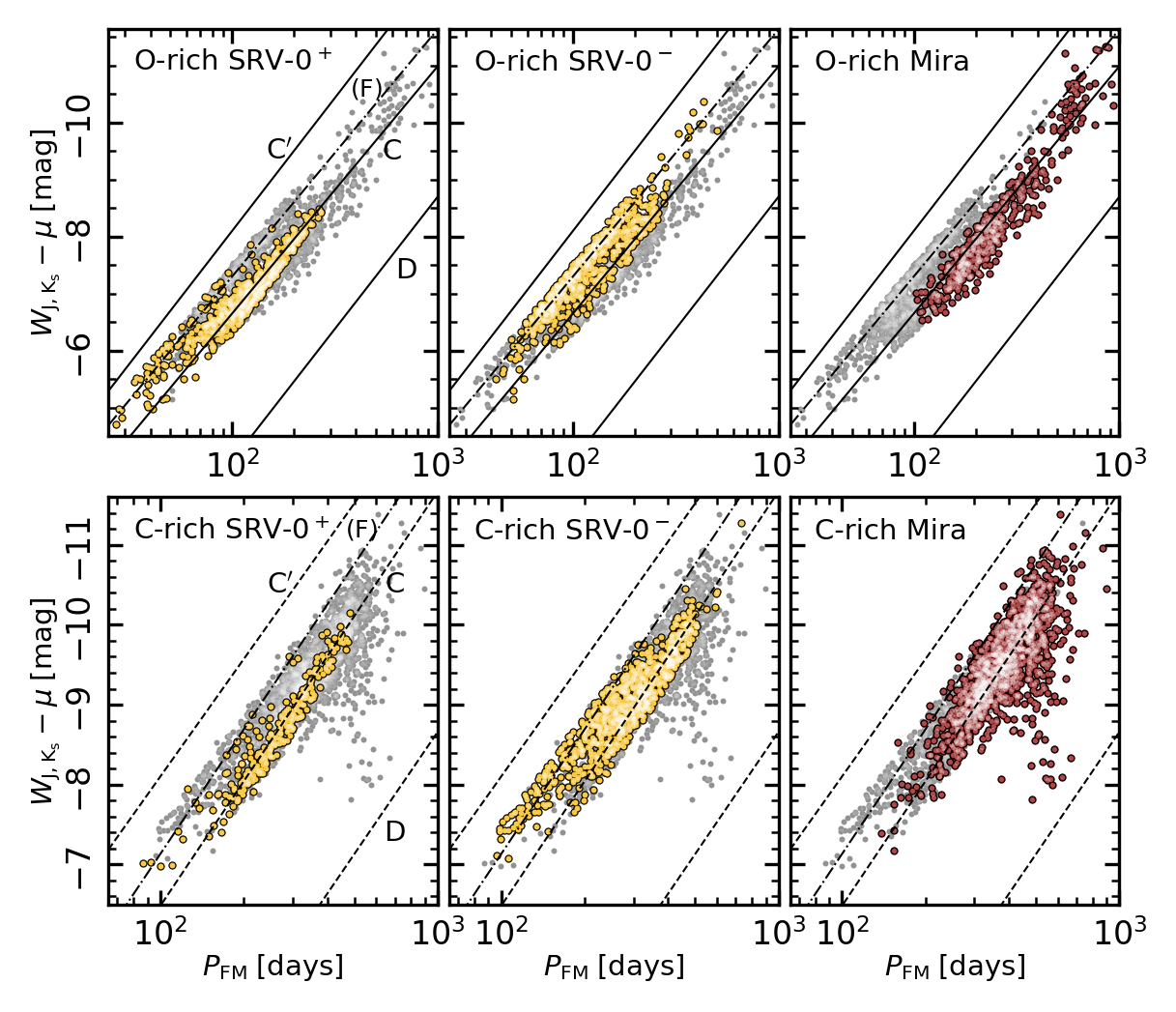}
    \caption{PLD of O-rich (top panels) and C-rich (bottom panels) LPVs pulsating only in the FM. Panels in the left and central columns show, respectively, SRV-0$^+$ and SRV-0$^-$ stars (cf. Fig.~\ref{fig:PAD_FM_SRVMira}). The former tend to align on sequence C, as Miras do (panels in the right columns), while the latter often lie on sequence F, between sequences \cprime\ and C. Solid and dashed lines are best-fit PLRs from \citet{Soszynski_etal_2007} (cf. Fig.~\ref{fig:PLD_SRVs_seqs}). As they do not provide a fit to sequence F, to indicate its approximate location, we used the best-fit equation for sequence C and offset it by $0.65$ mag towards brighter $\wjk$ (dash-dot line).}
     \label{fig:PLD_SRVsMiras_seqF}
\end{figure}

\begin{figure*}
    \centering
    \includegraphics[width=\hsize]{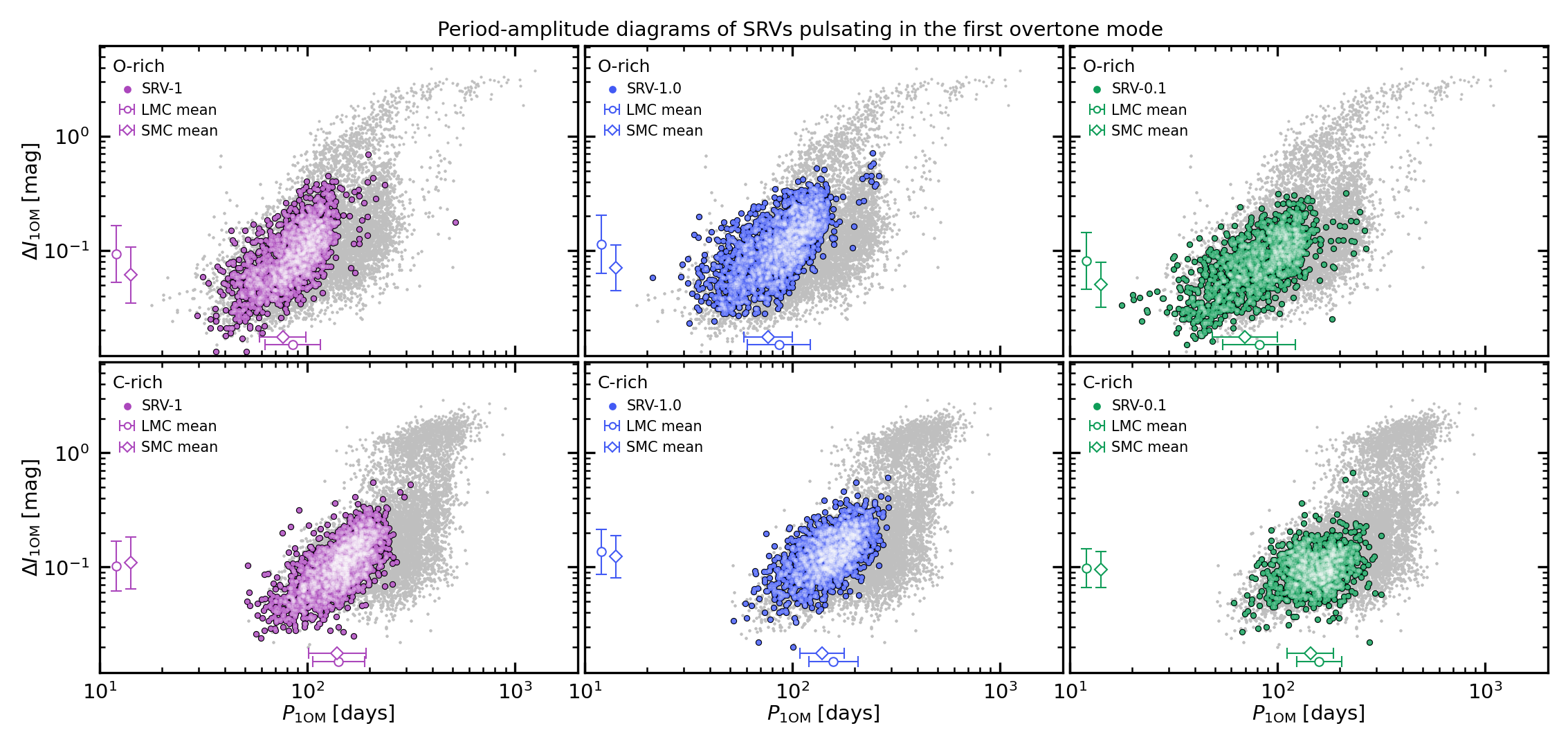}
    \caption{Similar to Fig.~\ref{fig:PAD_FM_SRVMira} but for the first overtone mode, showing sources assigned to groups SRV-1 (purple), SRV-1.0 (blue), and SRV-0.1 (green) shown in the panels in the left, central, and right columns, respectively.}
     \label{fig:PAD_1OM_SRVMira}
\end{figure*}

\begin{figure}
    \centering
    \includegraphics[width=.95\columnwidth]{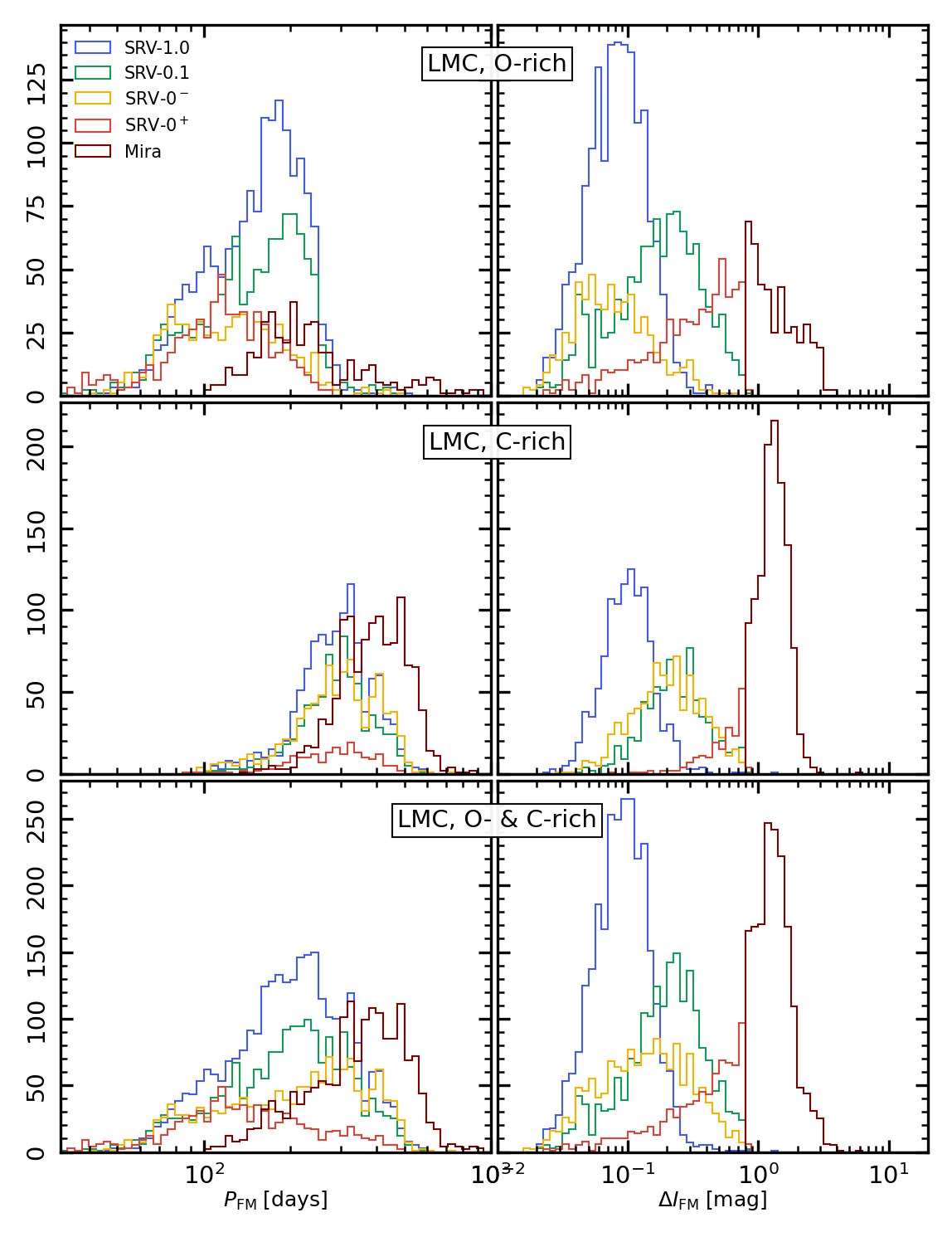}
    \caption{Period (panels on the left) and $I$-band amplitude (panels on the right) of the FM in stars in the LMC classified as SRV-1.0 (blue), SRV-0.1 (green), SRV-0$^-$ (orange), SRV-0$^+$ (red), and Miras (dark red). Panels in the top, middle, and bottom rows show O-rich stars, C-rich stars, and both, respectively.}
     \label{fig:hist_vtype_pafm}
\end{figure}

\begin{figure}
    \centering
    \includegraphics[width=.95\columnwidth]{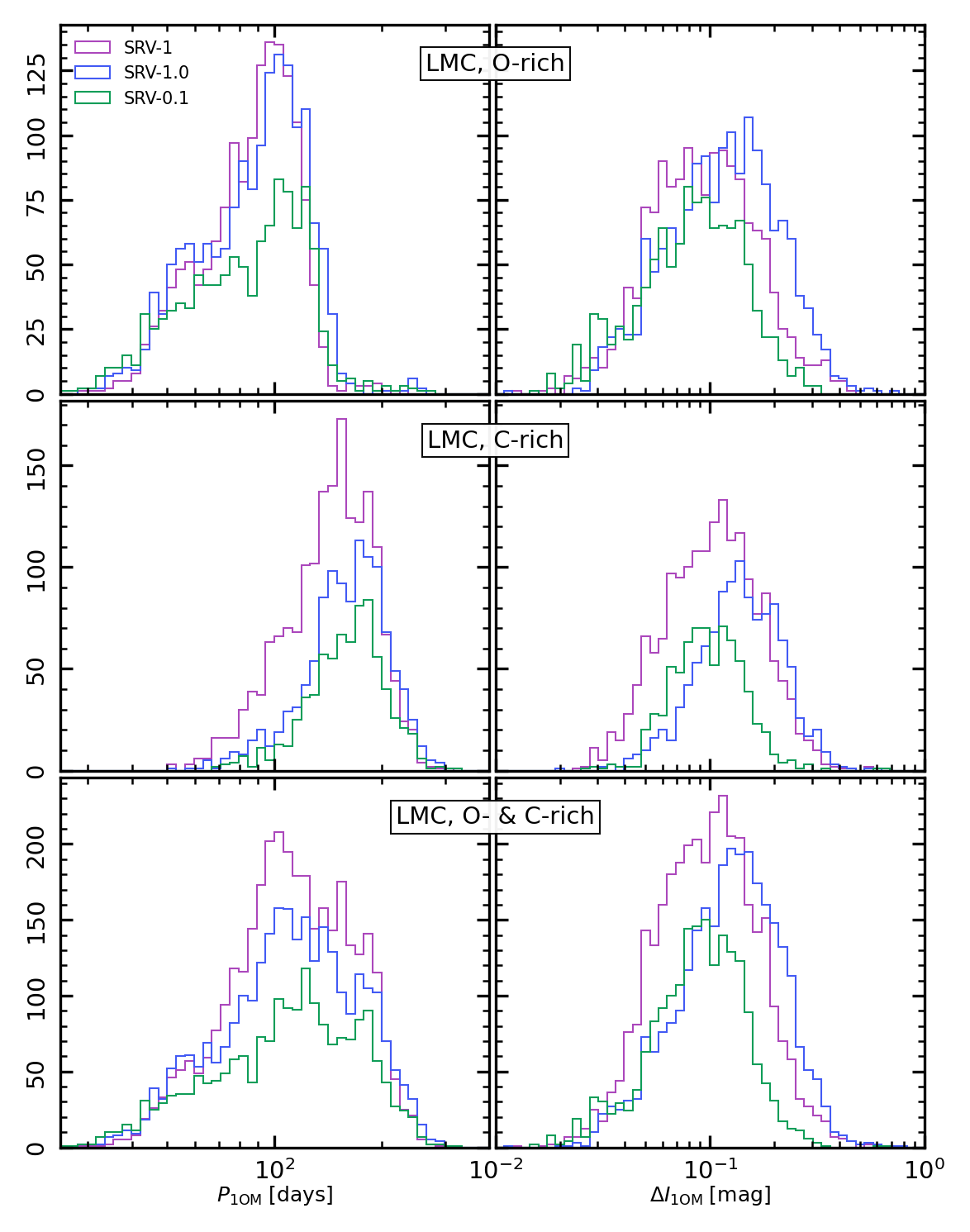}
    \caption{Similar to Fig.~\ref{fig:hist_vtype_pafm}, but for the 1OM, showing groups SRV-1, SRV-1.0, and SRV-0.1 in purple, blue, and green, respectively.}
     \label{fig:hist_vtype_pa1om}
\end{figure}

\begin{figure}
    \centering
    \includegraphics[width=.95\columnwidth]{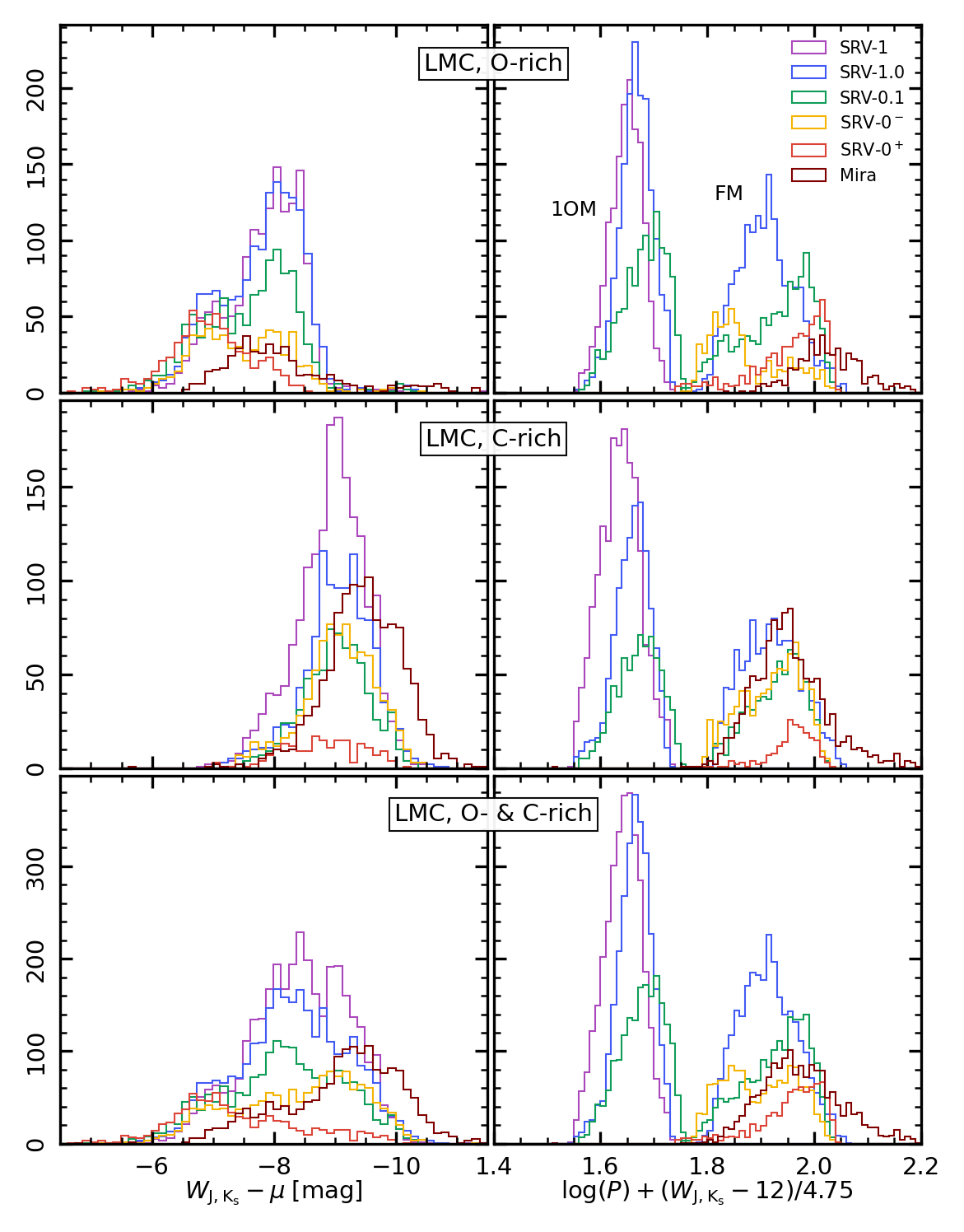}
    \caption{Similar to Fig.~\ref{fig:hist_vtype_pafm}, but showing the distributions of absolute $\wjk$ (panels on the left) and of $\Delta\log(P)$ (Eq.~\ref{eq:dlogp}).}
     \label{fig:hist_vtype_wjk_dlogp}
\end{figure}

The smaller-amplitude SRV-0$^-$ stars are comparable to SRV-1.0 stars (left-column panels of Fig.~\ref{fig:PAD_FM_SRVMira}) in terms of their distribution in the PAD while SRV-0$^+$ stars appear to be akin to Miras. Indeed, the SRV-0$^+$ stars seem to form a sequence of increasing period and amplitude that seamlessly merges into the distribution of Miras, although the amplitude of the latter tends to become constant towards the longest periods. The origin of the latter feature is unclear. A possible explanation is that non-linear processes intervene to limit the growth of physical amplitude of pulsation with luminosity \citep[as suggested by the results of][]{Trabucchi_etal_2021}, and that this is reflected in the behaviour of photometric amplitudes. Alternatively, it may be related to a weakening of the dependence of observed amplitude on temperature variations associated with a decreased sensitivity of molecular formation to temperature.

The traditional amplitude-based SRV--Mira separation is clearly inconsistent with these results, which is especially evident in the O-rich case. SRV-0.1 stars dominated by FM pulsation but showing also a 1OM period show similar albeit less expressed features. In particular, they do not show clear signs of two subgroups in the PAD, but do display a degree of continuity with the distribution of Miras.

The similarity between Miras and SRV-0$^+$ stars extends to the PLD (Fig.~\ref{fig:PLD_SRVsMiras_seqF}), where both are found to align on sequence C, although they tend to occupy different ranges of brightness. In contrast, the lower-amplitude SRV-0$^-$ stars are more frequently offset to the left of PL relation C, and are compatible with the location of the PL sequence labelled F \citep[first noticed by][]{Soszynski_etal_2005}, which is consistent with the findings of \citet{Soszynski_Wood_2013}. This trend is more evident in O-rich stars than in C-rich ones, and it is partially exhibited by SRV-0.1 stars as well, but not by SRV-1.0 stars. In contrast with Miras, the distribution of FM periods from all SRV groups covers the full range of sequences F and C, regardless of the existence of a clear separation.

\subsection{Comparison between SRV groups}
\label{ssec:ComparisonBetweenSRVGroups}

It is interesting to examine the differences between the PADs of distinct SRV groups, as well as of stars differing in chemical type or coming from different environments (i.e. located in the LMC or SMC). Let us now start with the PADs of 1OM pulsators shown in Fig.~\ref{fig:PAD_1OM_SRVMira}, where we have indicated the mean value and standard deviation of the period and amplitude distributions. The three groups SRV-1, SRV-1.0, and SRV-0.1 are very similar in terms of period distribution. On the other hand, there are some differences, albeit small, in the distributions of amplitudes. Indeed, SRV-1.0 stars tend to peak at larger 1OM amplitudes than SRV-1 and SRV-0.1 stars, while the latter two are comparable.

A similar analysis of Fig.~\ref{fig:PAD_FM_SRVMira} reveals that, in SRV-1.0 stars, the FM peaks at smaller amplitudes than in SRV-0.1 and SRV-0 stars, and covers a smaller range. The SRV-0.1 and SRV-0 groups on the other hand are comparable in terms of amplitude, while all three groups have similar period distributions. Figures~\ref{fig:hist_vtype_pafm} and~\ref{fig:hist_vtype_pa1om} show the histograms of periods and amplitudes for sources pulsating in the FM and 1OM, respectively, providing a more detailed view of their distributions (but limited for clarity to sources in the LMC).

It is also useful to examine the $\wjk$ luminosity function shown in Fig.~\ref{fig:hist_vtype_wjk_dlogp} (also limited to the LMC) together with the distribution of the quantity
\begin{equation}\label{eq:dlogp}
    \Delta\log(P) = \log(P) + (\wjk-12)/4.75 \,,
\end{equation}
which represents a horizontal coordinate across PL sequences in the $\log(P)$~--~$\wjk$ diagram. This is almost the same quantity as that defined by \citet{Wood_2015}, except this latter author adopted a slope of $-4.444$, which is representative of sequence A, while we use the value -4.75 because sequences \cprime \ and C are steeper than sequence A. In other words, Fig.~\ref{fig:hist_vtype_wjk_dlogp} displays the distribution of stars in terms of period ``along'' and ``across'' the PLRs in the PLD. The distributions of $\Delta\log(P)$ show that sources in groups SRV-1, SRV-1.0, and SRV-0.1 are gradually shifted towards the right edge of sequence \cprime. O-rich FM pulsators show a similar trend, as groups SRV-1.0, SRV-0.1, and SRV-0 show an increasing offset towards the long-period edge of sequence C, and O-rich Miras are the closest to the edge. This trend is also visible in C-rich stars for the 1OM, but not for the FM.

Such general differences between SRV groups, both for the 1OM and FM, do not depend on chemical type and apply to both the SMC and LMC. However, there are some trends that do depend on these properties. For instance, C-rich SRVs tend to have longer periods than O-rich stars. This is true for both the 1OM and the FM, and regardless of the SRV group they are in. Similarly, LPVs in the LMC tend to have longer periods than their SMC counterparts.

Furthermore, O-rich stars in the LMC tend to show significantly larger amplitudes than those in the SMC, but C-rich LPVs do not show such differences. Interestingly, within each group of SRVs, O-rich stars have almost the same average amplitude as C-rich stars (or just slightly higher for FM pulsators), but only in the LMC. In the SMC, C-rich SRVs always peak at larger amplitudes than O-rich ones. In other words, C-rich LPVs have the same mean amplitude in the LMC and SMC, which is comparable with the mean amplitude of O-rich LMC stars but is larger than that of O-rich sources in the SMC. These properties are consistent with the findings of \citet{Matsunaga_etal_2005} and \citet{Ita_etal_2021}.

\subsection{Colour--magnitude diagrams}
\label{ssec:ColourMagnitudeDiagrams}

\begin{figure*}
    \centering
    \includegraphics[width=\columnwidth]{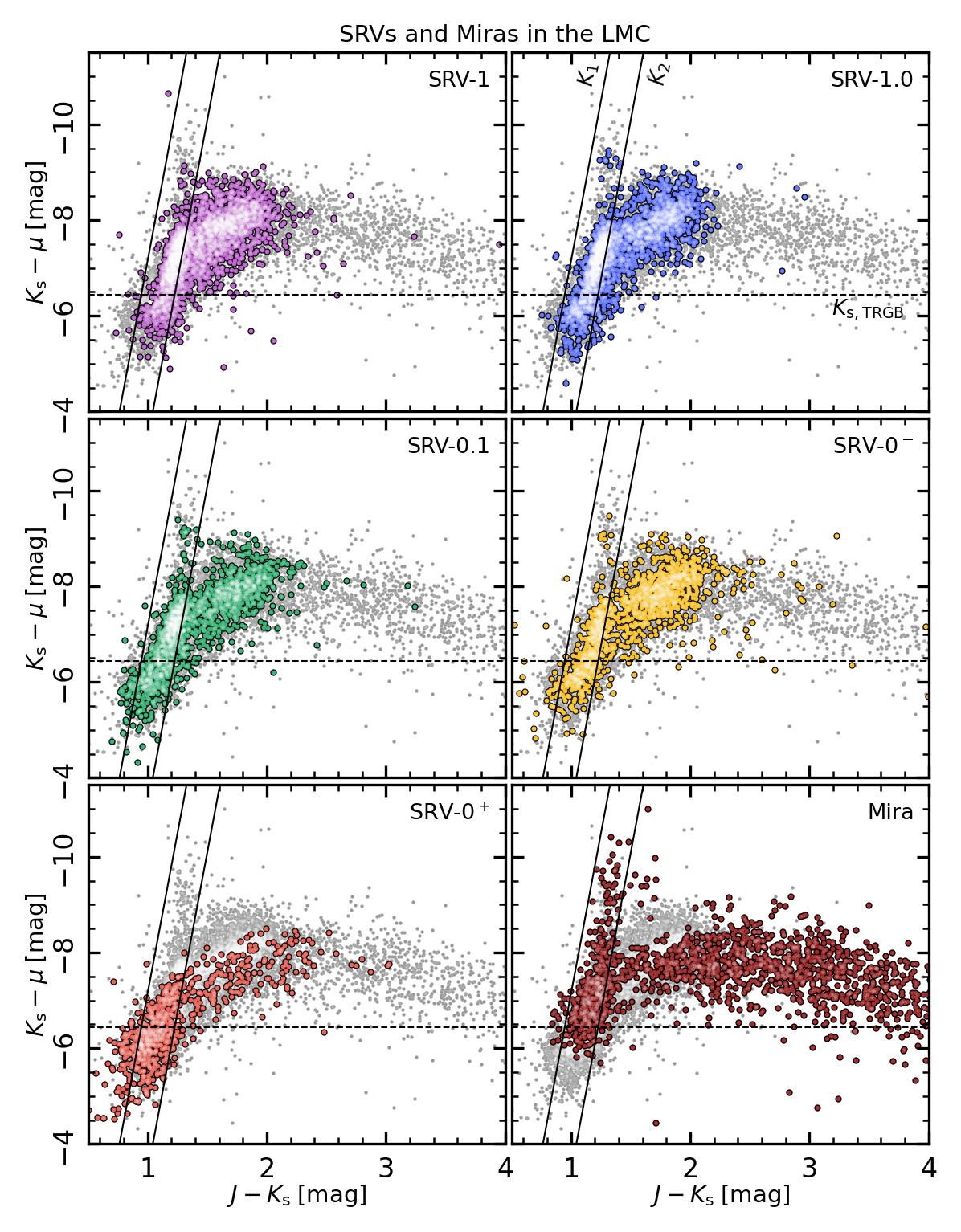}
    \includegraphics[width=\columnwidth]{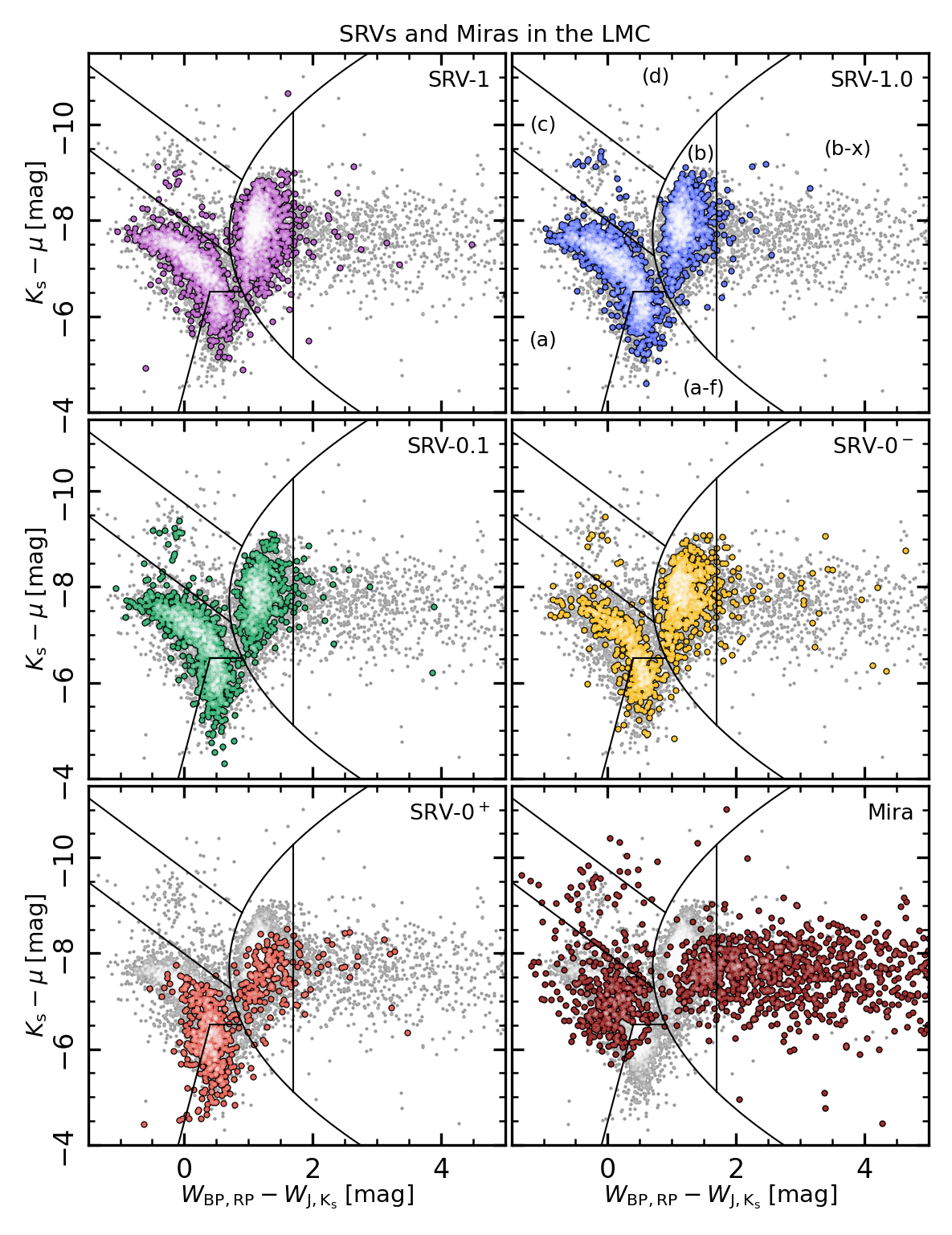}
    \caption{2MASS colour--magnitude diagrams (set of panels on the left) and \gtmd\ (set of panels on the right) of SRVs and Miras in the sample adopted here, limited to the LMC. Each panel in a given set corresponds to a different group of LPVs, according to the classification described in Sect.~\ref{ssec:PulsationModesIdentification}. The SRV-0 group is split into SRV-0$^-$ and SRV-0$^+$ (see Sect.~\ref{ssec:TheSRVMiraTransition}). The lines drawn to facilitate the comparison between 2MASS CMDs represent the RGB tip (dashed) and lines $K_1$, $K_2$ from \citet{Cioni_etal_2006}, while lines in the \gaia-2MASS diagrams are from \cite{Lebzelter_etal_2019}. We note that in both diagrams the horizontal range is truncated in order to focus on the region populated by SRVs, while Miras (and a few SRVs) extend as far as $J-\ks\simeq5$ mag and $\wbprp-\wjk\simeq9$ mag. \gaia\ EDR3 photometry is used to compute $\wbprp$.}
     \label{fig:JKsCMD_G2MD_SRVsMiras}
\end{figure*}

\begin{figure}
    \centering
    \includegraphics[width=\hsize]{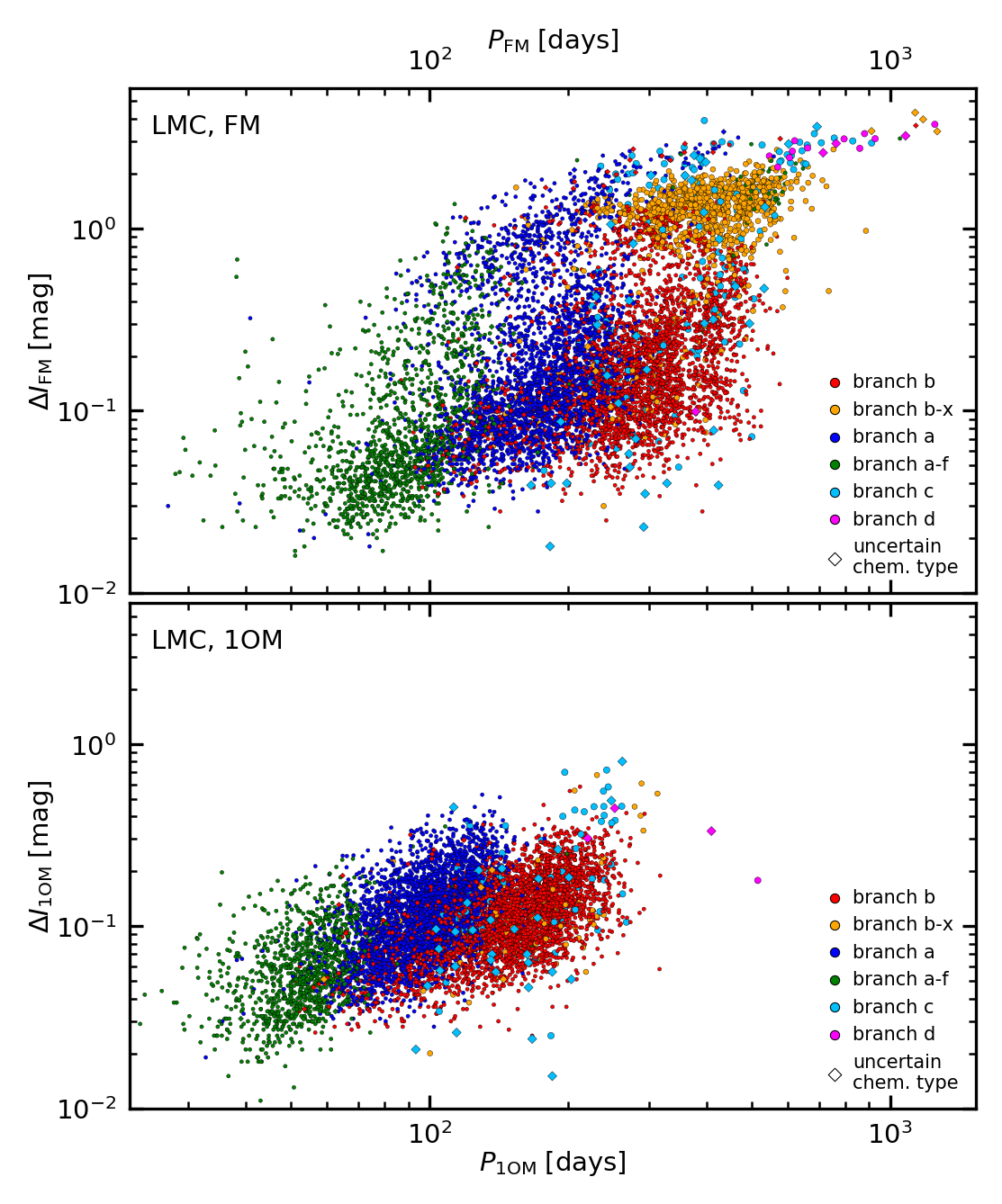}
    \caption{Period--amplitude diagram of the FM (top panel) and 1OM (bottom panel) of SRVs and Miras in the LMC, colour-coded by the branch of the G2MD they belong to. Sources with uncertain chemical type are identified by diamond symbols. The dashed line in the top panel has the same meaning as in Fig.~\ref{fig:PAD_FM_SRVMira}, i.e. it marks the assumed separation between O-rich SRV-0$^+$ and SRV-0$^-$ stars.}
     \label{fig:PAD_byG2MDtype_LMC}
\end{figure}

Figure~\ref{fig:JKsCMD_G2MD_SRVsMiras} shows the distribution of SRVs and Miras in the NIR colour--magnitude diagram (CMD) constructed with 2MASS photometry (panels on the left) and in the G2MD (panels on the right). We note that, for visual clarity, these panels are limited to stars in the LMC (the equivalent diagrams for the SMC are displayed in Appendix~\ref{asec:DiagramsForLPVsInTheSMC}) and sources with uncertain chemical type are excluded. Furthermore, SRV-0$^-$ and SRV-0$^+$ are displayed separately.
What is immediately clear from these diagrams is that all SRV groups display very similar trends, which are in stark contrast with the distribution of Miras, but SRV-0$^+$ stars show intermediate properties between SRVs and Miras.

Let us begin by considering C-rich LPVs, which show the largest differences between SRVs and Miras. These are the stars on branches $b$ and $b$-$x$ \citep[C-stars and extreme C-rich AGBs, respectively, see][]{Lebzelter_etal_2018} of the G2MD, and have $J-\ks\gtrsim1.5$ mag in the CMD. The most striking difference stems from C-rich Miras reaching much redder colours in the CMD and much larger values of $\wbprp-\wjk$, while C-rich SRVs are generally bluer than $J-\ks\simeq2.2$ mag, and are rarely found on branch $b$-$x$. Conversely, most C-rich Miras are on branch $b$-$x$, and only a small fraction of them are on branch $b$ or in the $J-\ks$ range where most C-rich SRVs are found ($1.5\lesssim J-\ks\,[{\rm mag}]\lesssim2.2$). Furthermore, C-rich SRVs peak at $\ks$ luminosities that are about $0.4$ mag brighter than C-Miras. C-rich SRV-0$^+$ are more similar to other SRV groups than to Miras, except for their average brightness, which is more similar to that of Miras (in the interval of $J-\ks$ where they overlap).

In the O-rich case, it is the Miras that have greater brightness on average. Only a few of them are found in the faint part of branch $a$ (labelled $a$-$f$ and occupied by faint, low-mass O-rich AGB stars), which hosts a substantial number of SRVs. Conversely, O-rich Miras are often found on branches $c$ and $d$, where only a few O-rich SRVs appear. It is likely that this difference is at least partially due to the bias introduced by the saturation of optically bright SRVs in OGLE, as pointed out in Sect.~\ref{ssec:Summary}, but the extent of this effect is unclear. Nonetheless, it is worth pointing out that sources on branches $c$ and $d$ also stand out in the PAD, and can be clearly recognised in the top panels of Fig.~\ref{fig:PAD_1OM_SRVMira} (the few SRVs with 1OM periods longer than $\sim150$ days), and of Fig.~\ref{fig:PAD_FM_SRVMira} (the SRVs with FM periods longer than $\sim350$ days and the Miras with periods longer than $\sim450$ days). These sources appear in the PAD of all SRV groups, regardless of the dominant pulsation mode.

The SRV-0$^+$ group represents an exception once again. Because of how it is defined in the PAD, it does not include any of the sources on branches $c$ or $d$. It is limited to branch $a$, but shows a distribution somewhat truncated to smaller $\ks$ and larger $\wbprp-\wjk$ than other SRVs. When compared with Miras, it is very tempting to interpret the former as the continuation at brighter $\ks$ of the SRV-0$^+$ distribution, which would be consistent with the evidence from the PAD.

Finally, in order to show the relation between the G2MD and the PAD, the latter is shown in Fig.~\ref{fig:PAD_byG2MDtype_LMC} with sources colour-coded by the branch of the G2MD they belong to, limited to LMC stars (the equivalent SMC diagram is shown in Appendix~\ref{asec:DiagramsForLPVsInTheSMC}). Different branches correspond to different regimes of initial mass, which shows an average increase with period in the PAD\footnote{
    We note that branch $c$ stars are expected to have systematically longer periods than branch $a$ stars (their lower mass counterparts) pulsating in the same mode. Instead, multiple stars classified as branch $c$ appear on the long-period side of the distribution of branch $a$ sources in the PAD. These stars most likely belong to branch $a$, but are located on its top edge, very close to the boundary between the two branches (cf. Fig.~\ref{fig:JKsCMD_G2MD_SRVsMiras}), and are probably erroneously attributed to branch $c$.
}. We note that such a clear differentiation would not arise unless periods were discriminated by their pulsation modes, as there is substantial overlap between FM and 1OM periods from different branches.

At the level of mass resolution allowed by the G2MD, there is no clear relation between mass and amplitude. On the other hand, we see that sources from branch $d$ (having the largest masses in the sample) and from branch $b$-$x$ (suffering from relatively large circumstellar extinction) are most often found in the large-amplitude region of the diagram. We did not find any other clear correlation of amplitude with any of the other photometric parameters that is not a reflection of an underlying correlation with period.

\section{Discussion}
\label{sec:Discussion}

Linear pulsation models show that overtone-mode pulsation becomes increasingly unstable as the stellar envelope grows in size. At the same time, the envelope cannot be too large, as it would fail the requirement for surface reflection of acoustic waves. Indeed, if the star expands beyond the limit allowing for surface reflection, the linear growth rate rapidly drops with increasing stellar radius. Therefore, throughout the AGB, the excitation degree of an overtone mode increases up to a maximum (at which point the mode is dominant), and then decreases until that mode becomes definitively stable.

As the maximum growth rate occurs earlier (at smaller radii) for modes of higher radial order compared with modes of lower order, a LPV goes through a sequence of dominant modes of decreasing radial order as it evolves, until it pulsates predominantly in the FM. At the same time, the number of excited modes decreases as they gradually become stable, but the surviving ones undergo a gradual increase in both period and growth rate, the latter representing a proxy for variability amplitude \citep[cf. fig~2 of][]{Trabucchi_etal_2017}.

Miras are (at least in most cases) mono-periodic FM pulsators with large amplitudes and relatively long periods, and so it is rather straightforward to identify them as the end-point of this pattern. With the aid of the classification we obtained in Sect.~\ref{sec:Classification}, we can also try to place SRVs into this scenario.
SRVs pulsating in the predominantly 1OM (SRV-1 and SRV-1.0) should be less evolved than SRVs dominated by FM pulsation (SRV-0.1 and SRV-0). Among the former two, SRV-1 should be (on average) the least evolved because the FM has not yet appeared. Similarly, the SRV-0 group should include objects that are, on average, more evolved than SRV-0.1 stars, because the 1OM has not yet become stable in the latter.

We therefore expect a LPV to go through the sequence of groups SRV-1$\to$SRV-1.0$\to$SRV-0.1$\to$SRV-0. This is supported by the fact that SRV-1 stars tend to have smaller 1OM amplitudes than SRV-1.0 stars, which have on average larger amplitudes than stars from group SRV-0.1. The latter should then be representative of stars in which the 1OM growth rate has already reached its maximum, and is now decreasing.

Moreover, stars in groups SRV-1, SRV-1.0, and SRV-0.1 are, in this order, gradually more shifted towards the right edge of sequence \cprime. This is in agreement with predictions from theory, according to which the period of a LPV increases across a PL sequence as the star evolves \citep[cf.][]{Trabucchi_etal_2019,Trabucchi_etal_2021}. This also means that the average stellar mass increases from SRV-0.1 through SRV-1.0 to SRV-1, because mass decreases from the short- to the long-period edges of a given PL sequence \citep{Wood_2015}. As it is unlikely that these stars have undergone significant mass loss, their current mass should be more or less representative of their initial mass.

We remark that there is likely a significant overlap in bass between these groups, and that this scenario suffers from a significant degeneracy with mass, upon which both the observed amplitudes and predicted growth rates depend. Indeed, the radius (or luminosity) at which a given overtone mode reaches maximum growth rate and becomes dominant depends on stellar mass. Moreover, the content of each group is determined by the initial mass function and star formation history for a given stellar population, and is affected by the fact that evolutionary timescales vary with stellar mass. The considerations made here should therefore be interpreted in a statistical sense, and are not expected to apply to every single case; they are nonetheless useful for the purpose of achieving a more clear understanding of LPVs.

Similar arguments can be put forward for FM pulsators, at least in part. There is indeed an increase in amplitude, as well as a shift towards the long-period edge of the PL sequence, when moving from SRV-1.0 to SRV-0.1. This trend extends further to SRV-0$^+$ stars, but the subgroup SRV-0$^-$ does not fit this pattern, in particular in the case of O-rich stars. Indeed, C-rich SRV-0$^-$ stars are offset towards the short-period edge of sequence C, to the point that they form another sequence, namely sequence F. We note that O-rich SRV-1.0 stars also extend into sequence F, but do not display an expressed division into two groups, either in the PAD or in the PLD. In contrast, C-rich SRV-0$^-$ stars appear very similar in their distribution to their SRV-0.1 counterparts.

Overall, the likenesses in variability properties between SRV-0$^+$ stars and Miras implies that stars belonging to this group represent the same type of object, that is, mono-periodic, large-amplitude FM pulsators, except the former are probably less evolved. The continuity observed in the PAD and PLD appears to be confirmed by the distributions in the CMD and in the G2MD. In the latter two, the position of SRV-0$^+$ stars suggests that, compared to Miras, they are less massive, because they are fainter; they also appear to be warmer, as they have less extreme colour (and $\wbprp-\wjk$) than Miras. The higher temperature is consistent with their lower photometric amplitude as this would hamper the formation of molecules, thus reducing the sources of visual opacity.

The group SRV-$0^-$ shows a distribution in the CMD and G2MD resembling that of `earlier' SRV groups, and populates more or less the same region of the PAD as the FM of SRV-1.0 stars. Compared to SRV-0$^+$ stars, they have smaller amplitudes at a given period, and follow a PLR offset towards shorter periods, suggesting their current masses are larger. These properties make it hard to include SRV-$0^-$ stars in the general scenario we put forward. For a better understanding, pulsation models capable of predicting photometric amplitudes of LPVs are required, which are currently unavailable due to the technical difficulties in dealing with the complex radiation-hydrodynamics of AGB atmospheres. We can, however, provide a number of possible explanations, which we list below and which will need to be verified by theory.

The simplest possible explanation for SRV-$0^-$ stars is that they are the less evolved version of SRV-$0^+$ stars, their FM still having to grow in period and amplitude. This is consistent with their position in the PLD, which suggests that they probably have larger current masses than SRV-$0^+$ stars. However, this does not explain the gap between the sequences F and C, or the splitting in the PAD, unless some relatively fast transition is invoked that produces an effect similar to the Hertzsprung gap in the Hertzsprung-Russell Diagram. Such a transition could be the shift from linear to non-linear pulsation, which would explain the similarities in the PAD with group SRV-1.0, in which the FM is the secondary mode and therefore is unlikely in the non-linear regime. This is also supported by the fact that the FM of SRV-$0^-$ stars span a similar interval of amplitudes as the 1OM (in other SRVs), whose properties are consistent with predictions from linear models.

We note that the 1OM of LPVs also splits into two sequences (B and \cprime), which was explained by \citet{Trabucchi_etal_2019} as a bias introduced by the appearance of large-amplitude LSP variability. The explanation provided by these latter authors is supported by the fact that LPVs with their primary period on sequence D have often another period between sequences B and \cprime. \citet{Soszynski_Wood_2013} observed a similar trend in that sequence F periods are often accompanied by LSP-like variability between sequences C and D, allowing for a possible extension of the interpretation of \citet{Trabucchi_etal_2017} to F-sequence stars.

An alternative possibility, based on a similar explanation, is that, when crossing between sequences F and C, the FM period is overshadowed by a resurfacing 1OM period rather than a LSP. This requires a mechanism able to reverse the normal pulsational evolution, and thermal pulses are the obvious candidates. Indeed, theory \citep{Trabucchi_etal_2019,Molnar_etal_2019} and observations \citep[e.g.][]{Szatmary_etal_2003} suggest that the envelope contraction following a He-shell flash can cause an LPV to return to a previous configuration, and stabilised overtone modes can become excited and dominant once again. In this scenario, SRV-0 stars would temporarily turn into SRV-0.1 or SRV-1.0 stars again, which would explain their likeness in the PAD.

Finally, the possibility should be mentioned that SRV-$0^-$ stars are actually the natural direct progenitors of Miras, and that the SRV-$0^+$ group actually comes from a different production channel. For instance, the latter might have suffered more intense mass loss \citep[they indeed show higher infrared excess, cf. Fig.~1][]{McDonald_Trabucchi_2019} as a consequence of binary evolution. However, there is no indication from OGLE data of any variability suggestive of binarity.

Regardless of the origin of the SRV-0 splitting, our analysis confirms that the traditional separation between Miras and SRVs is arbitrary and should be replaced by a criterion taking into account both period and amplitude, as well as the number of unstable pulsation modes \citep[cf.][]{Kiss_etal_2000}. Adopting FM-dominated LPVs rather than Miras alone and thus being more inclusive towards SRVs would be physically more justified. This would also be profitable for distance determinations of stellar populations. Even the most conservative extension, limited to SRVs pulsating only in the fundamental mode and with relatively large amplitudes (i.e. the SRV-0$^+$ group), would represent an overall $\sim50\%$ increase in available sources. If only O-rich LPVs are considered, as is often done because of their well-behaved PLR, the increase in sample size would be of $\simeq$140\%. It should be noted, however, that excluding sources from group SRV-0$^-$ would introduce some degree of arbitrariness in the selection because of how the two subtypes of SRV-0 stars are defined.

Semi-regular variables that pulsate simultaneously in the FM and 1OM provide a means to achieve an additional, substantial boost in the number of potential LPV distance indicators, but are more challenging because of their multi-periodicity. The main limitations from a practical point of view are connected with the availability of photometric time-series of appropriate quality for their characterisation. These need to cover at least a few years of observations to be relatively densely sampled and to have photometric precision at the level of $\gtrsim0.01$. It is entirely realistic to expect these requirements to be satisfied by current and upcoming observational surveys, many of which are explicitly targeted to time-domain astronomy.

As an example, we point out that SRV-0$^+$ stars have $I$-band amplitudes typically larger than $\sim0.2$ mag, which is essentially the amplitude limit of the \gaia\ DR2 catalogue of candidate LPVs \citep{Mowlavi_etal_2018}. This means that at least one extensive full-sky catalogue for these stars is already available. Moreover, the single period provided for \gaia\ DR2 LPVs is enough for such stars, because they are mono-periodic. The upcoming \gaia\ DR3 is expected to reach smaller amplitudes thereby giving access to sources compatible with the other SRV groups.

The \gaia\ mission is of additional interest because it delivers multi-band photometric time-series, similarly to the Legacy Survey of Time and Space \citep[LSST, ][]{Ivezic_LSST_2109} of the upcoming Vera Rubin Observatory. Such surveys will allow the study of LPV amplitudes at various wavelengths.

An important issue that will need to be addressed for using multi-periodic SRVs as distance indicators is the larger scatter of their PLRs. This is also relevant for SRV-0 stars, because they effectively follow two distinct PLRs, but in principle they can be identified by their different amplitude regime. In this sense, additional variability properties besides the dominant period are expected to be crucial in overcoming this difficulty. In particular, it is worth keeping in mind that, while multi-periodicity may be challenging from the point of view of observations and processing, each period beyond the primary carries additional physical information to be exploited. A detailed investigation of the role of multi-periodicity, as well as variability amplitudes, for distance determinations will be the subject of a forthcoming paper.

The analysis of the PAD also revealed trends associated with the chemical type and the environment. C-rich LPVs have, on average, longer periods than O-rich ones because they are brighter\footnote{
    Massive O-rich AGB stars are in fact brighter than C-rich AGBs, but they are more rare, and therefore play a small role in determining average properties, at least at the metallicities of the Magellanic Clouds.
}. With the exception of a small offset, this applies to both the LMC and SMC. Furthermore, C-rich LPVs have, on average, larger amplitude than O-rich stars, but while the difference is small in the LMC, it is considerable in the SMC.

This environmental effect is most likely due to the average metallicity difference between the LMC and SMC. In the latter, the lower metallicity favours the production of C-stars, meaning that there are less O-rich stars than in the LMC as they have already become C-rich. This leads to the loss of O-rich sources near the O- to C-rich transition, i.e. O-rich LPVs with relatively high brightness and large amplitude. As a result, the mean amplitude of O-rich stars is shifted towards smaller values.

However, this cannot fully explain the observed trends; for example, according to this scenario, the periods should be affected comparably, which is not the case. Therefore, there is probably a more direct effect of metallicity at play. This is not surprising, as the photometric variability of LPVs is largely determined by the spectral absorption bands of molecules, whose formation is controlled by temperature as well as metallicity. Pulsation models able to simulate LPV light curves are needed to better understand these trends.

If confirmed, these metallicity-related trends would be very interesting. On the one hand, the PAD would have some potential as a diagnostic tool to examine the chemical composition of LPV populations. On the other hand, the fact that the variability of SRVs carries a trace of their metallicity can be useful if they are to be applied as distance indicators for stellar populations with a different chemical composition from that of the stars used as calibrators. In these respects, expanding the analysis of the PAD with observations in various photometric filters, possibly with the help of model predictions, is expected to be very profitable.

\section{Conclusions}
\label{sec:Conclusions}

We analysed LPVs in the Magellanic Clouds, focusing on the similarities and differences between Miras and SRVs in order to prepare the ground for studying the potential of the latter as distance indicators.

Starting from the OGLE-III catalogues of LPVs in the Magellanic Clouds, we assembled a data set consisting of Miras and SRVs that pulsate in the FM or 1OM. We classified them by number of active pulsation modes and corresponding radial order, and identified four groups of mono- and multi-periodic SRVs that fit a general evolutionary scenario for the pulsation of luminous red giants. The classified data set is made publicly accessible.
The main results of this research can be summarised as follows.\begin{itemize}
    \item   Semi-regular variables pulsating only in the FM and with relatively large amplitude form a continuous distribution with Miras in the period--amplitude and period--luminosity diagrams, which is further evidence that the traditional amplitude-based classification of LPVs should be revised.
    \item   Being mono-periodic, these SRVs do not pose significantly higher difficulties on practical grounds with respect to Miras, and their adoption as standard candles would more than double the size of the sample of LPVs available for this purpose.
    \item   Lower-amplitude and multi-periodic SRVs also show similarities to Miras, and are substantially more numerous. While promising, they involve specific challenges as they do not follow a single PLR and display a somewhat larger scatter in the period--luminosity diagram. Strategies to address this will be presented in the next paper of this series.
    \item   We highlight the diagnostic power of the PAD in the investigation of long-period variables, as it is effectively independent of distance and reddening \citep[although possibly affected by crowding; e.g.][]{Riess_etal_2020}. The comparative analysis of samples with different chemical types and environmental metallicities suggests that it might have some potential for studying the chemistry of stellar populations, at least in a statistical sense.
\end{itemize}

\begin{acknowledgements}
We are grateful to Prof. Igor Soszynski for the constructive comments that helped improving this paper.
M.T. and N.M. acknowledge the support provided by the Swiss National Science Foundation through grant Nr. 188697.
This publication makes use of data from the \mbox{OGLE-III} Catalogue of Variable Stars.
This publication makes use of data products from the Two Micron All Sky Survey, which is a joint project of the University of Massachusetts and the Infrared Processing and Analysis Center/California Institute of Technology, funded by the National Aeronautics and Space Administration and the National Science Foundation. This work has made use of data from the European Space Agency (ESA) mission {\it Gaia} (\url{https://www.cosmos.esa.int/gaia}), processed by the {\it Gaia} Data Processing and Analysis Consortium (DPAC, \url{https://www.cosmos.esa.int/web/gaia/dpac/consortium}). Funding for the DPAC has been provided by national institutions, in particular the institutions participating in the {\it Gaia} Multilateral Agreement.
This publication makes use of the following free/open source software and/or libraries: the Starlink Tables Infrastructure Library \citep[STILTS and Topcat][]{Taylor_2006}; IPython \citep{ipython} and Jupyter \citep{jupyter} notebooks; the \textsc{Python} libraries \textsc{NumPy} \citep{numpy2020}, \textsc{SciPy} \citep{SciPy}, \textsc{matplotlib} \citep[a \textsc{Python} library for publication quality graphics][]{matplotlib}, and \textsc{Astropy} \citep[a community-developed core \textsc{Python} package for Astronomy][]{astropy2018}.
\end{acknowledgements}

\bibliographystyle{aa}
\bibliography{references}

\appendix

\section{Diagrams for LPVs in the SMC}
\label{asec:DiagramsForLPVsInTheSMC}

\begin{figure}
    \centering
    \includegraphics[width=\hsize]{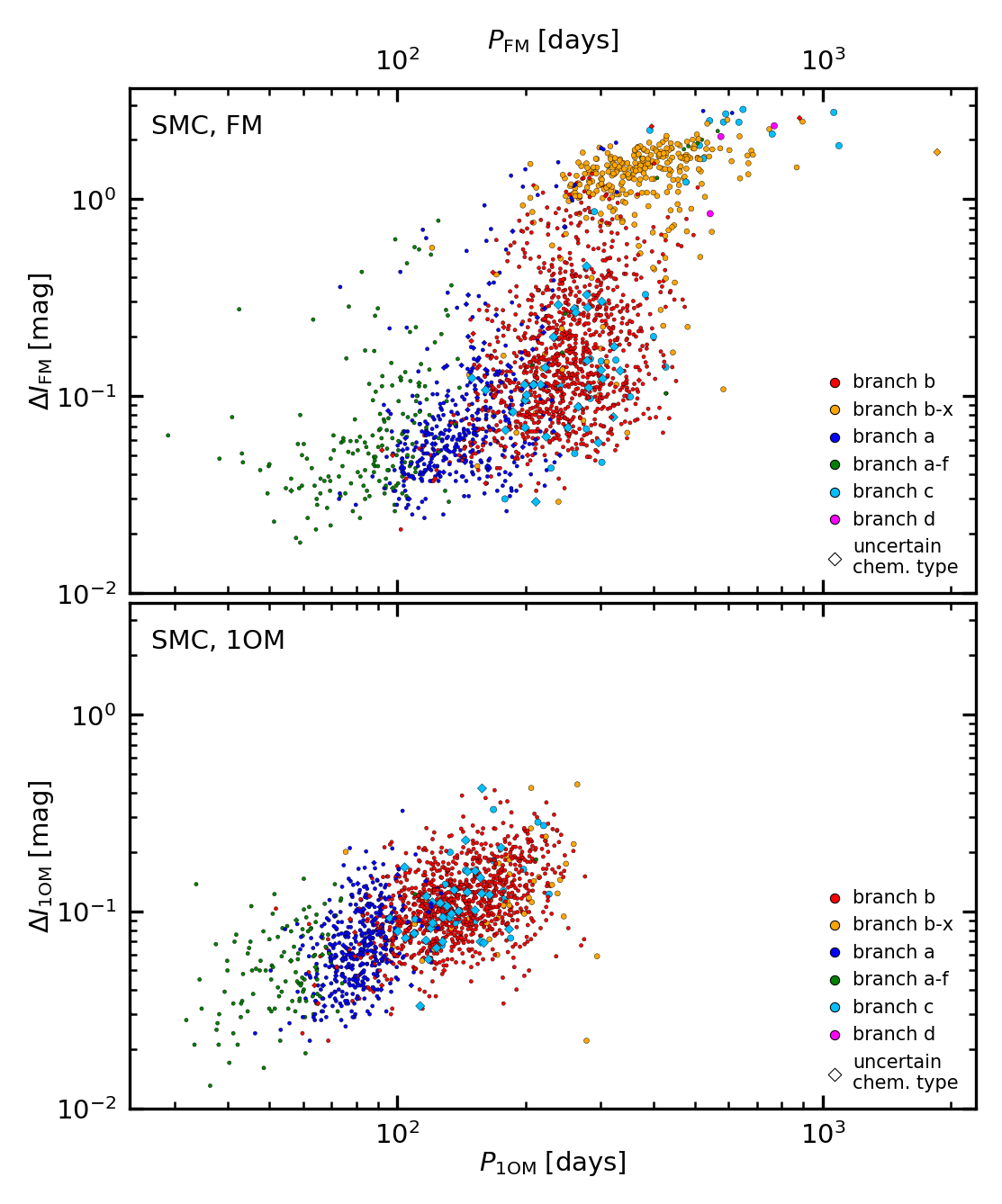}
    \caption{Similar to Fig.~\ref{fig:PAD_byG2MDtype_LMC}, but for sources in the SMC.}
     \label{fig:PAD_byG2MDtype_SMC}
\end{figure}

\begin{figure*}
    \centering
    \includegraphics[width=\columnwidth]{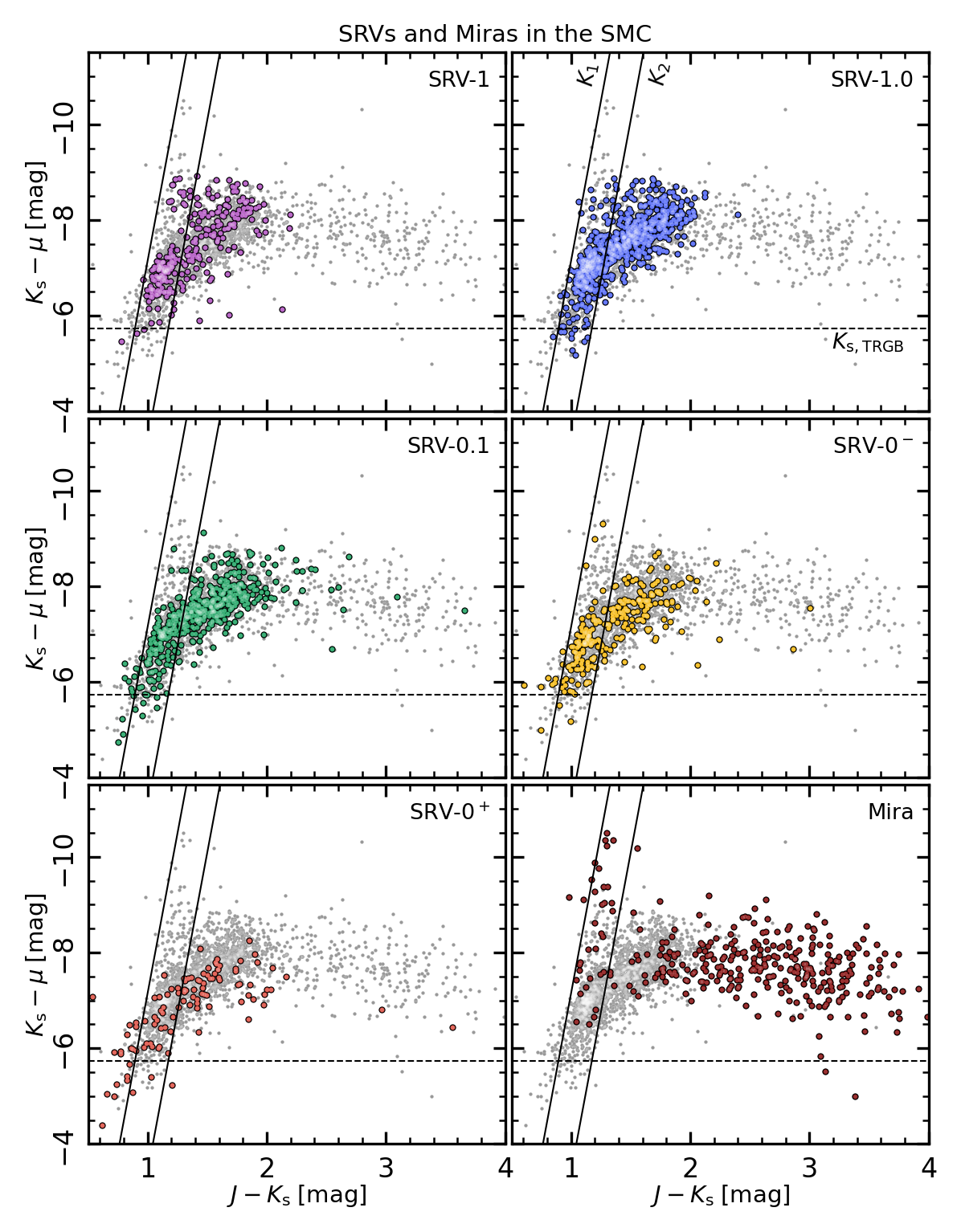}
    \includegraphics[width=\columnwidth]{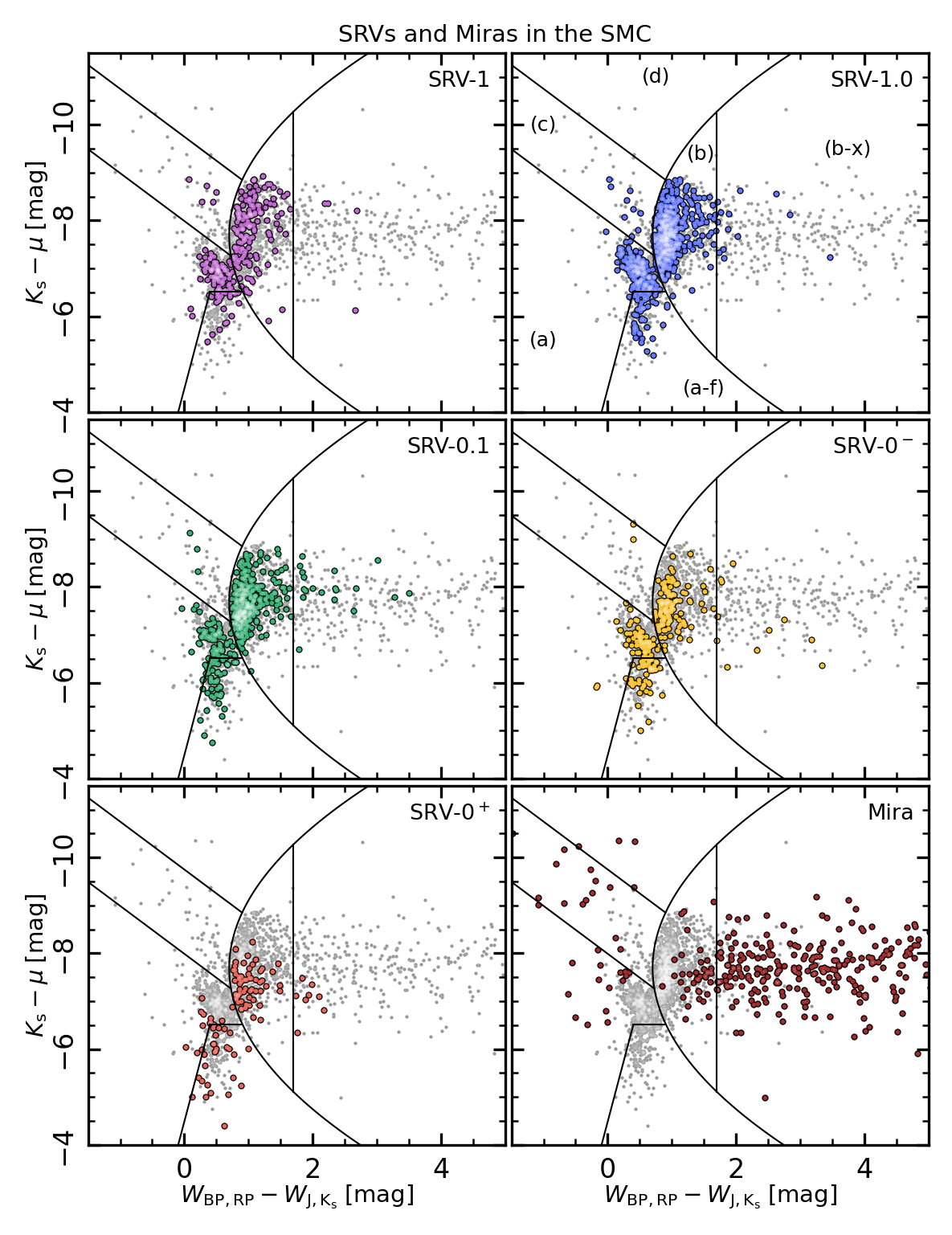}
    \caption{Similar to Fig.~\ref{fig:JKsCMD_G2MD_SRVsMiras}, but for sources in the SMC.}
     \label{fig:JKsCMD_G2MD_SRVsMiras_SMC}
\end{figure*}

We display here a few additional figures showing the distribution of SRVs and Miras in the SMC. Figure~\ref{fig:PAD_byG2MDtype_SMC} shows the PAD, while Fig.~\ref{fig:JKsCMD_G2MD_SRVsMiras_SMC} shows the NIR CMD and the G2MD. These are to be compared, respectively, with Figs.~\ref{fig:PAD_byG2MDtype_LMC} and~\ref{fig:JKsCMD_G2MD_SRVsMiras} in Sect.~\ref{ssec:ColourMagnitudeDiagrams}.

\end{document}